\documentclass[11pt,a4paper]{article}
\usepackage[margin=1in]{geometry}
\usepackage[T1]{fontenc}
\usepackage{amsmath,amssymb,mathtools,amsthm,graphicx,booktabs}
\usepackage{newtxtext}
\usepackage[subscriptcorrection]{newtxmath}
\usepackage{microtype,subcaption,xcolor}
\usepackage[authoryear,round]{natbib}
\usepackage[colorlinks=true,allcolors=blue!45!black]{hyperref}
\newcommand{\R}{\mathbb R}
\newcommand{\Pp}{\mathbb P}
\newcommand{\E}{\mathbb E}
\newcommand{\diam}{\operatorname{diam}}

\newtheorem{theorem}{Theorem}
\newtheorem{proposition}[theorem]{Proposition}
\title{Inference on Agreement Among Analytical Specifications}
\author{Debashis Chatterjee\\\small Department of Statistics, Visva-Bharati, Santiniketan 731235, India\\\small\href{mailto:debashis.chatterjee@visva-bharati.ac.in}{debashis.chatterjee@visva-bharati.ac.in}}
\date{}
\hypersetup{pdftitle={Inference on Agreement Among Analytical Specifications},pdfauthor={Debashis Chatterjee}}
\begin{document}
\maketitle

\begin{abstract}
When several analyses of the same data give different estimates, it is often unclear how much disagreement remains at the population level. We study the smallest interval containing a prescribed weight of population effects and construct simultaneous confidence bounds for its width. The two endpoints have different mathematical structures. The lower endpoint is determined by shortest paths and a weighted maximum-flow problem, whereas the upper endpoint must account for competing subsets within a common feasible configuration. We characterize the sharp upper endpoint via linear programs and present a case in which a simpler upper bound is arbitrarily loose relative to it. The bands require neither a reference analysis nor a unique optimal subset. Simulations illustrate the advantage of paired contrasts and the failure of selected pointwise intervals. In experimental earnings data, the nominal band supports agreement on at least half of the outcome definitions within 1,000 dollars, although the estimated full range is 1,395 dollars.
\end{abstract}
\par\medskip

\section{Introduction}
Suppose that several researchers analyze the same data to answer a common scientific question. Some estimates agree closely, whereas others differ substantially. Does this mean that the underlying effects disagree, or that the estimates are simply noisy? And if agreement is possible only among a subset of the analyses, how large a subset can the data support? These questions arise naturally in the many-analyst studies of \citet{silberzahn2018} and \citet{aczel2026}, the latter of which involves reanalyses of 100 social and behavioral studies. They also arise when a single researcher varies outcome definitions, adjustment sets, or modeling assumptions.

Two features of this problem call for a statistical treatment. First, analyses of common observations are rarely independent. An uncertain estimate of each of two effects may still give a precise estimate of their difference if the sampling errors move together. Second, the structure of disagreement matters. Five effects near zero and one far away describe a different situation from six evenly spaced effects with the same overall range. A measure of total spread cannot distinguish these situations. Nor is a collection of pairwise decisions sufficient: the subset that appears most coherent is itself selected using the data.

Multiverse analysis \citep{steegen2016} and specification curves \citep{simonsohn2020} provide ways to display variation across defensible choices. Joint inference and post-selection inference are already important parts of this literature \citep{girardi2024}. Our question is complementary: rather than test a common null across specifications, we seek a confidence statement about the amount of population-level agreement, even when the targets differ. Following \citet{del2021}, we distinguish choices that change the scientific target from those that change only its estimation. Our framework requires that targets be comparable on a common scale, not that they be identical by definition. It also makes no claim that agreement establishes correctness.

We measure agreement by the length of the shortest interval containing a given weight of the population effects. The interval functional is related to the concentration summaries \citep{einmahl2010}; the difference here lies in the observation scheme. The interval must contain unknown, jointly estimated population quantities, rather than observations drawn from a distribution of effects. Simultaneous inference for linear contrasts is classical \citep{tukey1949,hothorn2008}. We ask what these contrast constraints imply about the shortest interval, allowing the best subset to vary over the confidence region.

The key is an asymmetry. To make the smallest feasible interval as short as possible, it suffices to find one admissible subset and one feasible configuration. To make that interval long, a single configuration must keep \emph{every} admissible subset dispersed. We show that the first problem reduces to a partial order and a weighted maximum flow problem, while the second reduces to linear programs over possible orderings. Even when all pairwise upper bounds are identical, treating the subsets separately can be arbitrarily inefficient. These geometric results separate calibration from computation and give simultaneous statements over all retained weights and tolerances. They remain applicable at ties, where ordinary smooth approximations to the optimal interval can fail.

\section{The population quantity}
Let $m\ge2$ be the number of prespecified analyses, indexed by $[m]=\{1,\ldots,m\}$, with population targets $\theta=(\theta_1,\ldots,\theta_m)^\top\in\R^m$. Let $w_i>0$ be fixed weights with $\sum_i w_i=1$, chosen independently of the inference data. For a subset $S\subseteq[m]$, write $w(S)=\sum_{i\in S}w_i$ and $\diam_S(x)=\max_{i,j\in S}(x_j-x_i)$. For $0<q\le1$, set $\mathcal S_q=\{S:w(S)\ge q\}$ and define
\begin{equation}\label{eq:rho}
 R_q(\theta)=\min_{S\in\mathcal S_q}\diam_S(\theta).
\end{equation}
Thus $R_q$ is the least tolerance within which analyses of total weight at least $q$ can agree. Equivalently, for tolerance $t\ge0$, the greatest weight that can be retained is
\begin{equation}\label{eq:mass}
 A_t(\theta)=\max_{a\in\R}\sum_{i=1}^m w_i\mathbf1\{a\le\theta_i\le a+t\},
\end{equation}
where $\mathbf1\{\cdot\}$ is an indicator. The equivalence $R_q\le t$ if and only if $q\le A_t$ is exact; $1-A_t$ is the smallest weight that must be set aside at tolerance $t$. Both profiles are monotone and have finitely many breakpoints. They do not change under a common translation, and $R_q$ scales with the absolute value of a common scale factor. Splitting an analysis into identical copies with the same total weight leaves the population profiles unchanged. With equal weights, $R_q$ is the shortest interval containing $\lceil mq\rceil$ entries; sorting and a two-pointer scan compute it in $O(m\log m)$ time for any $q$.

For instance, the equally weighted vectors $(0,0,0,0,0,1)$ and $(0,0.2,0.4,0.6,0.8,1)$ have the same range, but at weight $5/6$ their interval lengths are zero and 0.8. Reporting $R_1$ alone hides this distinction. A weighted profile describes not only how far apart the effects are, but also how much disagreement must be tolerated to include additional analyses. We do not interpret the weights as probabilities that an analysis is correct.

\section{Sharp projection of a contrast region}
Suppose that for some $0<\alpha<1$ finite random bounds $b_{ij}$ satisfy
\begin{equation}\label{eq:event}
 \Pp_\theta(\theta_j-\theta_i\le b_{ij}\text{ for all }i\ne j)\ge1-\alpha.
\end{equation}
Define the region $C=\{x\in\R^m:x_1=0,\ x_j-x_i\le b_{ij},\ i\ne j\}$. Fixing the first coordinate removes a common translation; it does not privilege the first analysis in the definition of agreement. If $C$ is empty, we report a vacuous band $[0,\infty)$; otherwise it is compact. In a directed graph with vertices $[m]$, give edge $i\to j$ length $b_{ij}$. Write $d_{ij}$ for shortest-path distances, with $d_{ii}=0$. Feasibility rules out negative cycles; all distances are finite. The closure inequalities $x_j-x_i\le d_{ij}$ are equivalent to the original constraints and can be found by the algorithm of \citet{floyd1962}. This closure is important: an indirect comparison through a third analysis may be more precise than the direct comparison.

\begin{theorem}[Lower endpoint]\label{thm:lower}
If $C$ is nonempty, the sharp lower endpoint is
\begin{equation}\label{eq:lower}
 L_q=\min_{x\in C}R_q(x)=\min_{S\in\mathcal S_q}\max_{i,j\in S}(-d_{ij}).
\end{equation}
For $t\ge0$, put $i\prec_t j$ if $d_{ji}<-t$. This is a strict partial order. Form a network with source $s$, sink $v$, and two copies $i_L,i_R$ of each vertex. Edges $s\to i_L$ and $i_R\to v$ have capacity $w_i$; edges $i_L\to j_R$ have capacity 2 whenever $i\prec_t j$. If $F_t$ is the maximum flow value, then
\begin{equation}\label{eq:flow}
 \max_{x\in C} A_t(x)=1-F_t,\qquad L_q=\min\{t\ge0:1-F_t\ge q\}.
\end{equation}
\end{theorem}
The graph for the lower endpoint is not an arbitrary compatibility graph. Indeed, $i\prec_t j$ means that $x_j-x_i>t$ throughout $C$. Such forced separation is transitive. An eligible subset is coherent if and only if it is an antichain, a set with no two comparable members. The weighted antichain formula is an application of the chain--antichain duality of \citet{dilworth1950} and maximum-flow/minimum-cut duality \citep{ford1956}. Formula (\ref{eq:flow}) also gives a direct computation of the sharp upper endpoint for $A_t$. For $L_q$, binary search over the sorted candidate values $0$ and $\max(0,-d_{ij},-d_{ji})$ requires $O(\log m)$ flow evaluations after closure. Thus, the lower projection has a polynomial-time algorithm, without enumerating subsets or solving a general maximum-clique problem. The proof in the Appendix explains why pairwise compatibility is sufficient here.

For the upper endpoint, compatible pairwise configurations need not be the same. For a permutation $\pi$ of $[m]$, define the order cell $C_\pi=C\cap\{x_{\pi(1)}\le\cdots\le x_{\pi(m)}\}$. For each starting index $a$ with $\sum_{k=a}^m w_{\pi(k)}\ge q$, let $h_\pi(a)$ be the first index $b\ge a$ with $\sum_{k=a}^b w_{\pi(k)}\ge q$.
\begin{theorem}[Upper endpoint]\label{thm:upper}
For each nonempty $C_\pi$, let $u_\pi$ be the solution value of the linear program
\begin{equation}\label{eq:lp}
 \max_{x\in C_\pi,\ z\ge0}z,\qquad
 z\le x_{\pi(h_\pi(a))}-x_{\pi(a)}\quad\text{for every eligible }a.
\end{equation}
Then $U_q=\max_{x\in C}R_q(x)=\max_{\pi:C_\pi\ne\varnothing}u_\pi$, and
\begin{equation}\label{eq:cheap}
 U_q\le\widetilde U_q=\min_{S\in\mathcal S_q}\max_{i,j\in S}d_{ij}.
\end{equation}
At full weight, $U_1=\widetilde U_1=\max_{i,j}d_{ij}$.
\end{theorem}
For any fixed ordering, only the shortest eligible block starting at each index matters. This turns the maximum of a minimum of linear functions into (\ref{eq:lp}). The cells cover $C$, including ties. Formula (\ref{eq:cheap}) avoids permutation enumeration, though its clique problem is not in general the same partial-order problem as (\ref{eq:flow}). We make no polynomial-time claim for the sharp upper projection.

The distinction is not merely technical. Suppose all weights are equal and $C$ consists of the anchored configurations with diameter at most $D>0$. For an integer $2\le k\le m$ and $q=k/m$, put $p=\lfloor(m-1)/(k-1)\rfloor$. Then
\begin{equation}\label{eq:gap}
 U_q=D/p,\qquad \widetilde U_q=D.
\end{equation}
To see this, sort any configuration. The $p$ consecutive intervals joining ranks $1, k, 2k-1,\ldots,1+p(k-1)$ each contain $k$ points, so their lengths are at least $R_q(x)$ and sum to at most $D$. Equality is attained by the ordered entries $D\lfloor(i-1)/(k-1)\rfloor/p$ for $i=1,\ldots,m$. Every block of $k$ entries then spans $D/p$. In particular, for $k=2$ the ratio of the separate-subset bound to the sharp bound is $m-1$. Three equally weighted effects already give $D/2$ rather than $D$. Sharpness concerns the realized confidence region, not an assertion of optimality over all confidence procedures.

\section{Simultaneous inference}
\begin{theorem}[Coverage and inversion]\label{thm:coverage}
Under (\ref{eq:event}), with probability at least $1-\alpha$, for all $0<q\le1$ and $t\ge0$ simultaneously,
\[
 L_q\le R_q(\theta)\le U_q,\qquad
 \sup\{q:U_q\le t\}\le A_t(\theta)\le\sup\{q:L_q\le t\}.
\]
The suprema are over $(0,1]$, with the empty set having supremum zero. Replacing $U_q$ by $\widetilde U_q$ preserves validity.
\end{theorem}
On the event in (\ref{eq:event}), the centered true vector lies in $C$, which proves the first statement by translation invariance. The second follows from (\ref{eq:mass}). Importantly, $U_q\le t$ certifies that a coherent subset of weight at least $q$ exists; $L_q>t$ rules out its existence. In the intervening case, the data do not resolve the question. To certify a particular subset from $C$, its maximum closed upper contrast bound must not exceed $t$. Thus, the conclusion need not identify a common subset across feasible configurations. The simultaneous statement allows inspection of $q$ and $t$ after computing the band, not data-driven changes to the universe or its weights.

The contrast region can be calibrated independently of the projection. Suppose $\widehat\theta-\theta\sim N_m(0,V)$ with known covariance $V$. Write $s_{ij}^2=V_{ii}+V_{jj}-2V_{ij}$ and let $c$ be the $(1-\alpha)$ quantile of $\max_{i<j:s_{ij}>0}|G_j-G_i|/s_{ij}$ for $G\sim N_m(0,V)$. Use $b_{ij}=\widehat\theta_j-\widehat\theta_i+cs_{ij}$; zero-variance contrasts are exact, and if all contrast variances are zero, use $c=0$. The Bonferroni choice $c=\Phi^{-1}\{1-\alpha/[m(m-1)]\}$ is conservative, with $\Phi$ the standard normal distribution function. With estimated covariance and fixed $m$, plug-in calibration is asymptotically valid under a centered joint Gaussian limit, consistent covariance estimation on the corresponding scale, positive limiting contrast variances, and continuity of the limiting maximum distribution. Resampling, where justified, must act on common observational units.

A finite-sample construction is available for $\theta_i=\E f_i(Z)$ with estimator $\widehat\theta_i=n^{-1}\sum_{k=1}^n f_i(Z_k)$, where $Z_1,\ldots,Z_n$ are independent and identically distributed and $f_i$ are fixed measurable functions. If the paired difference $f_j(Z)-f_i(Z)$ lies in a known interval of length $B_{ij}=B_{ji}$, Hoeffding's inequality \citep{hoeffding1963} and a union bound yield
\[
 b_{ij}=\widehat\theta_j-\widehat\theta_i+B_{ij}\left\{\frac{\log(m(m-1)/\alpha)}{2n}\right\}^{1/2}.
\]
The range of a paired difference can be far smaller than the sum of the marginal ranges. For independent arms of sizes $n_0,n_1$, with the same known range in each arm, replace $1/n$ by $1/n_0+1/n_1$; details are in Section~S2 of the Supplementary Material.
\begin{proposition}[Contraction]\label{prop:rate}
For $b_{ij}=\widehat\theta_j-\widehat\theta_i+e_{ij}$, where $e_{ij}=e_{ji}\ge0$, put $e_*=\max_{i\ne j}e_{ij}$ and $\widehat R_q=R_q(\widehat\theta)$. For all $q$,
\[
 \widehat R_q-e_*\le L_q\le\widehat R_q\le U_q\le\widetilde U_q\le\widehat R_q+e_*.
\]
Both bands have width at most $2e_*$. In the bounded-score case the width is of order $B_*\{\log(m/\alpha)/n\}^{1/2}$, with $B_*=\max_{i\ne j}B_{ij}$.
\end{proposition}
The contraction result requires no unique minimizer. To understand why this matters, at complete population agreement the scaled plug-in profile converges, under a joint root-$n$ Gaussian limit, to the shortest-interval functional of the Gaussian vector itself, generally not a Gaussian variable. At a tie between optimal subsets, the limit is a minimum of active directional range derivatives. Section S1 presents the general directional derivative and its proof within the framework of \citet{fang2019}. The contrast projection avoids the need to estimate these active sets for calibration. The width bound is not a claim of minimax optimality.

\section{Numerical illustrations}
\subsection{Monte Carlo experiment}
We generated independent Gaussian observation vectors of dimension six, with means all zero (tie), four equally spaced values from zero to 0.04 followed by 0.45 and 0.52 (core), or six equally spaced values from zero to 0.5 (gradient). The marginal standard deviations range evenly from 0.75 to 1.25. The correlations are independent, equicorrelated at 0.9, or $0.7vv^\top+0.3I_m$, where $v$ has half positive and half negative unit entries and $I_m$ is the identity matrix. Sample sizes are 100 and 500. Two ten-specification core scenarios and four unequally weighted bounded-score scenarios complete the 24-scenario design, each with 1,000 replications. See Section~S3 for full settings and all scenario-specific results.

The main simulation uses $[L_q,\widetilde U_q]$ with nominal 95\% calibration. Paired Gaussian maxT uses known covariance and 120,000 Gaussian calibration draws per scenario. We compare it with paired Bonferroni, its estimated-covariance version, a Bonferroni marginal box, working independence, and a selected pointwise interval based on the empirically shortest block; the last two are diagnostics, not procedures with a simultaneous guarantee. Paired Hoeffding is used in the bounded-score cases. Coverage is evaluated at every attainable weight breakpoint, and widths are also averaged over these breakpoints. At coverage 0.95, the Monte Carlo standard error is 0.007.
\begin{figure}[t]
\centering\includegraphics[width=\linewidth]{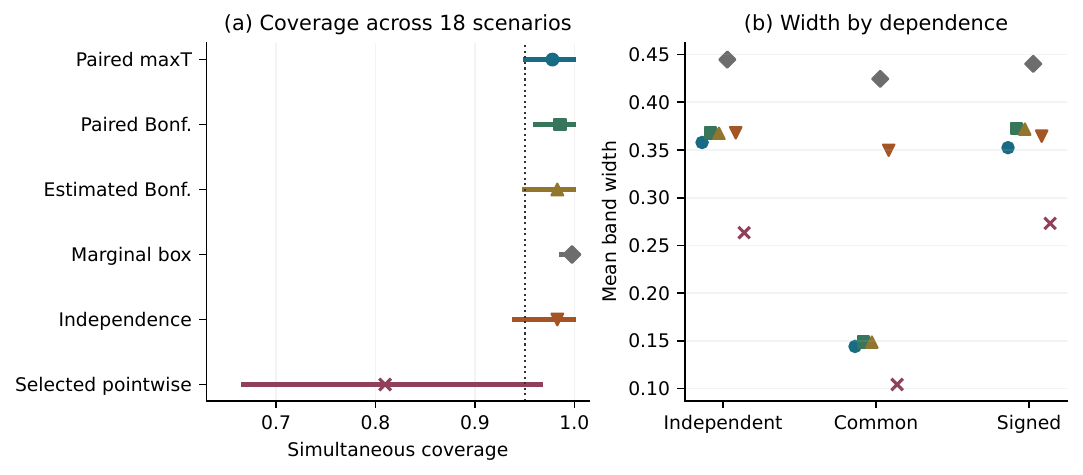}
\caption{Six-specification Gaussian experiments: (a) mean simultaneous coverage and range over 18 scenarios; (b) mean bandwidth by dependence structure, averaged over mean configurations and sample sizes. The dotted line denotes 0.95. Segments in (a) indicate scenario ranges, not confidence intervals. Symbols in both panels, in the order of the labels in (a), are circle, square, upward triangle, diamond, downward triangle, and cross.}\label{fig:sim}
\small\noindent\textit{Alt text:} Coverage is near or above nominal for paired methods; selected pointwise intervals cover less. Under common dependence, the paired bands are much shorter than marginal or independence bands.
\end{figure}
Across Gaussian scenarios, maxT simultaneous profile coverage ranged from 0.950 to 1.000, with 0.950--0.960 under ties. Figure~\ref{fig:sim} shows the role of dependence: under common noise, the mean widths were 0.144 for maxT, 0.350 for working independence, and 0.425 for the marginal box. Under signed dependence, working independence gave coverage of 0.944 and 0.939 in the two tie scenarios. Selected pointwise coverage fell to 0.667 in Gaussian cases and 0.657 with bounded scores. Hoeffding covered all 4,000 bounded-score replications, at the expense of wider bands than paired normal calibration. Projection itself can also be conservative: in one signed-gradient scenario with a sample size of 100, contrast-region coverage was 0.934, but profile coverage was 0.995. A vector outside the confidence region may have the same profile as a vector inside it.

Independent linear programs verified 216 lower endpoints over 108 asymmetric regions to within $1.12\times10^{-16}$. In a separate four-specification experiment (900 replications), the sharp upper endpoint was smaller than the conservative endpoint at half weight in 48.1\% of replications (mean gap 0.021), while the gap vanished at full weight, as predicted by Theorem~\ref{thm:upper}. The two bands had identical observed simultaneous coverage across all six scenarios (0.953--1.000). Section S4 reports timings, additional gap results, and numerical checks. The maximum-flow characterization is verified separately by exhaustive subset comparisons in Section~S1; the main Monte Carlo uses the equivalent original subset calculation.

\subsection{Sensitivity of experimental earnings to upper-tail treatment}
The National Supported Work experiment is a useful example of sensitivity to the definition of outcomes \citep{lalonde1986,dehejia1999,dehejia2002}. We use the Dehejia--Wahba subsample (185 treated and 260 control participants) as distributed in R datasets \citep{nswmixtape,nswarchive}. Let $Y$ denote 1978 earnings in thousands of historical US dollars, and let $T$ denote the treatment indicator. We define $f_i(y)=\min(y,c_i)$ with $c_i\in\{5,10,15,20,30,\infty\}$, taking $f_i(y)=y$ at infinity, and estimate $\theta_i=\E[f_i(Y)\mid T=1]-\E[f_i(Y)\mid T=0]$. These are distinct outcome estimands, not alternative estimators of the uncapped mean effect. For finite caps $a<b$, the change in the target equals the integral from $a$ to $b$ of the difference in the two earnings survival functions. Hence changing the cap probes the role of the upper tail, rather than choosing a statistically convenient estimate.

Under an independent two-sample superpopulation model conditional on arm counts, estimate $\theta$ by $\bar f_1-\bar f_0$ with covariance estimate $\widehat V=S_1/n_1+S_0/n_0$, where $\bar f_a,S_a,n_a$ are the sample mean vector, covariance and size in arm $a\in\{0,1\}$. We use equal weights and plug-in Gaussian maxT (120,000 draws); stratified participant bootstrap (10,000 draws), paired Bonferroni, and capped-only Hoeffding serve as sensitivity analyses.
\begin{table}[t]
\centering
\caption{Experimental earnings: individual estimates and reconciliation profile, in thousands of historical US dollars. SE denotes marginal standard error; profile bounds use nominal 95\% plug-in maxT. Rows do not associate a cap with a particular subset.}\label{tab:nsw}
\small
\begin{tabular}{lrr@{\hspace{1.1em}}crrr}
\multicolumn{3}{c}{Outcome definition}&\multicolumn{4}{c}{Reconciliation profile}\\
Cap & Estimate & SE & $q$ & $\widehat R_q$ & $L_q$ & $U_q$\\[2pt]
5 &0.399&0.214&$1/6$&0.000&0.000&0.000\\
10&0.855&0.382&$2/6$&0.185&0.000&0.479\\
15&1.141&0.481&$3/6$&0.427&0.000&0.977\\
20&1.367&0.540&$4/6$&0.653&0.000&1.493\\
30&1.609&0.596&$5/6$&0.939&0.000&2.015\\
Uncapped&1.794&0.671&$1$&1.395&0.047&2.784
\end{tabular}
\end{table}
Table~\ref{tab:nsw} shows that the estimated effect rises with the cap. The estimated correlations of the specification estimates range from 0.695 to 0.988. For caps 15 and 20 the paired standard error of the difference is 0.099, compared with 0.723 if covariance is ignored. At a tolerance of one, the estimated profile suggests agreement among five of six outcomes; the simultaneous band certifies at least three, not five. The marginal box has upper endpoint 2.575 at half weight, compared with 0.977 for paired maxT. At full weight, the band is $[0.047,2.784]$, compared with bootstrap $[0.024,2.813]$; paired Bonferroni includes zero, so evidence against exact agreement is calibration-sensitive. The sharp and conservative upper bounds coincide here. In the capped-only universe, the full-weight Hoeffding band is $[0,5.372]$ and the plug-in band $[0.090,2.329]$. Section S5 presents the profiles, inverse profiles, and weight and calibration sensitivity; separate calibrations do not protect different universes jointly. A large weight on a single specification also makes small-weight agreement trivial, so weights require scientific justification.

\section{Discussion}
The interval length describes agreement rather than the truth of a scientific conclusion. All analyses may agree and still share a bias; a certified core may also consist of different subsets at different feasible configurations. The application concerns estimand sensitivity under a superpopulation model, not exact finite-population randomization inference or a complete causal multiverse. For fitted nonlinear specifications, the calibration must be justified afresh. A useful open problem is to avoid enumerating order cells for large universes while retaining the sharpness of the upper endpoint. The lower endpoint has a much simpler structure, and the gap in (\ref{eq:gap}) cautions against interchanging the choice of subset and the maximization of uncertainty.

\section*{Declaration of the use of AI-assisted technologies}
During the preparation of this work, OpenAI was used to assist with Python programming and grammatical and syntactic editing. 

\section*{Acknowledgement}
The author is also an Adjunct Fellow at the S. N. Bose National Center for Basic Sciences, Kolkata, India, where part of the research was done.
\section*{Supplementary material}
The Supplementary Material includes full proofs, directional limit theory, computational checks, calibration details, extended Monte Carlo results and further analysis of experimental earnings (Sections S1--S6).
\section*{Data and code availability}
The exact public input is the \href{https://vincentarelbundock.github.io/Rdatasets/csv/causaldata/nsw_mixtape.csv}{Rdatasets NSW CSV}; the original data archive is cited in the references. The Python code and notebook are available under the MIT license at \url{https://github.com/debashisdotchatterjee/Reconciliation_Profile_Stat2026}. The additional flow audit accompanies the source files.

\section*{Appendix: Proofs}
\noindent\textit{Proof of Theorem~\ref{thm:lower}.}
Fix $S$ and $t\ge0$. If $\diam_S(x)\le t$ for some $x\in C$, then $d_{ij}\ge-t$ for all $i,j\in S$. Conversely, add directed edges of length $t$ between all distinct vertices of $S$. A cycle with $k$ new edges consists of these edges and $k$ old paths between vertices of $S$, each of length at least $-t$; the cycle cannot be negative. Old cycles are nonnegative. A shortest-path potential from an auxiliary source yields a feasible vector that can be centered. Minimizing in $t$ and $S$ proves (\ref{eq:lower}). The triangle inequality implies that $\prec_t$ is transitive, so a subset is feasible at tolerance $t$ exactly when it is an antichain. The flow identity follows by chain decomposition of the capacitated flow and the minimum-cut characterization, proved in Section~S1.

\noindent\textit{Proof of Theorem~\ref{thm:upper}.}
In a fixed cell, any eligible subset with extreme ranks $ a$ and $b$ satisfies $b\ge h_\pi(a)$; the block from $a$ to $h_\pi(a)$ is itself eligible. Thus its shortest range is the minimum of the block ranges, proving (\ref{eq:lp}). For fixed $S$, $R_q(x)\le\diam_S(x)\le\max_{i,j\in S}d_{ij}$, giving (\ref{eq:cheap}). For $q=1$, the potential $x_k=d_{ik}-d_{i1}$ is feasible by the triangle inequality and attains $x_j-x_i=d_{ij}$. Maximize over pairs.

\noindent\textit{Proof of Proposition~\ref{prop:rate}.}
The centered estimate lies in $C$. All feasible contrasts differ from their empirical values by at most $e_*$. Finite maxima and minima preserve this bound. Since $d_{ij}\le b_{ij}$, evaluate (\ref{eq:cheap}) at an empirically minimizing subset. The claimed inequalities and the rate follow. Further details, including the proof of the directional limit, appear in Section~S1 of the Supplementary Material.

\clearpage
\section*{Supplementary Material}
\phantomsection
\addcontentsline{toc}{section}{Supplementary Material}
\setcounter{section}{0}
\setcounter{subsection}{0}
\setcounter{equation}{0}
\setcounter{theorem}{0}
\setcounter{figure}{0}
\setcounter{table}{0}
\renewcommand{\thesection}{S\arabic{section}}
\renewcommand{\theequation}{S\arabic{equation}}
\renewcommand{\thetheorem}{S\arabic{theorem}}
\renewcommand{\thefigure}{S\arabic{figure}}
\renewcommand{\thetable}{S\arabic{table}}
\renewcommand{\theHsection}{supp.\arabic{section}}
\renewcommand{\theHsubsection}{\theHsection.\arabic{subsection}}
\renewcommand{\theHequation}{supp.\arabic{equation}}
\renewcommand{\theHtheorem}{supp.\arabic{theorem}}
\renewcommand{\theHfigure}{supp.\arabic{figure}}
\renewcommand{\theHtable}{supp.\arabic{table}}

\noindent This material provides complete proofs, additional theoretical results, calibration details, numerical verification, and extended analyses of the simulation and earnings data.
\section{Theoretical results and proofs}\label{supp:sec:proofs}
\subsection{Definitions}
Let $m\ge2$ be a fixed number of prespecified analyses, indexed by $[m]=\{1,\ldots,m\}$. Their population targets are $\theta=(\theta_1,\ldots,\theta_m)^\top\in\R^m$, expressed on a common scientifically interpretable scale. They need not be identical estimands. Let $w_i>0$ be fixed weights with $\sum_i w_i=1$, chosen independently of the inference data; write $w(S)=\sum_{i\in S}w_i$. For $0<q\le1$, let $\mathcal S_q=\{S\subseteq[m]:w(S)\ge q\}$. Define
\begin{equation}\label{supp:eq:rho}
 R_q(\theta)=\min_{S\in\mathcal S_q}\bigl(\max_{i\in S}\theta_i-\min_{i\in S}\theta_i\bigr).
\end{equation}
This is the minimum interval width needed to retain weight at least $q$; it is not a radius measured from a predetermined reference. It is nondecreasing in $q$, translation invariant, and homogeneous under multiplication by a scalar in absolute value. Equal weights recover the shortest interval containing at least $\lceil mq\rceil$ entries. Sorting the entries and scanning contiguous blocks computes the target in $O(m\log m)$ time, including sorting: for each starting index, use the first ending index whose cumulative weight reaches $q$, advancing the ending index monotonically.

For tolerance $t\ge0$, an equivalent formulation is
\begin{equation}\label{supp:eq:mass}
 A_t(\theta)=\max_{a\in\R}\sum_{i=1}^m w_i\mathbf1\{a\le\theta_i\le a+t\}.
\end{equation}
The maximum is attained since there are finitely many possible retained sets. Then $A_t\ge q$ if and only if $R_q\le t$. In particular, $1-A_t$ is the smallest weight that must be discarded to achieve tolerance $t$. Both profiles are determined by finitely many subset-weight or range breakpoints; an entire confidence profile therefore need not be approximated on an arbitrary grid. Weighting can prevent redundant variants of a pipeline from dominating: splitting a specification into identical copies, with weights that sum to the original weight, leaves both targets unchanged. This is an identity at the population level, not a recommendation to count near-duplicates as identical.

\subsection{Projection results}
Suppose that real-valued bounds $b_{ij}$ ($i\ne j$) satisfy
\begin{equation}\label{supp:eq:event}
 \Pp_\theta\{\theta_j-\theta_i\le b_{ij}\ \text{for all }i\ne j\}\ge1-\alpha,
\end{equation}
where $0<\alpha<1$. Define $C=\{x\in\R^m:x_1=0,\ x_j-x_i\le b_{ij}\ \forall i\ne j\}$. The anchor removes an irrelevant translation. If $C$ is empty, report the vacuous band $[0,\infty)$ for every $q$. This rule preserves coverage. Since bounds are finite and all ordered pairs are present, a nonempty $C$ is compact. In its directed graph, edge $i\to j$ has length $b_{ij}$; denote shortest-path distances by $d_{ij}$, setting $d_{ii}=0$. Negative edges are permitted, but feasibility excludes negative cycles.

\begin{theorem}[Sharp lower projection]\label{supp:thm:lower}
For every $q\in(0,1]$ and every nonempty $C$,
\begin{equation}\label{supp:eq:lower}
 L_q:=\min_{x\in C}R_q(x)
 =\min_{S\in\mathcal S_q}\max_{i,j\in S}(-d_{ij}).
\end{equation}
For any $r\ge0$, $L_q\le r$ if and only if the undirected graph joining $i$ and $j$ when both $d_{ij}\ge-r$ and $d_{ji}\ge-r$ has a clique of weight at least $q$. Equivalently, the graph is the incomparability graph of the strict partial order $i\prec_r j$ if $d_{ji}<-r$. The maximum weight of an antichain equals $1-F_r$, where $F_r$ is the network flow defined and proved in Section~\ref{supp:sec:flowproof}. In particular $L_q=\min\{r\ge0:1-F_r\ge q\}$.
\end{theorem}
The lower bound is sharp relative to the realized confidence region: there is a feasible configuration attaining it. In particular, closing the contrast bounds under shortest paths can eliminate apparently compatible but jointly impossible configurations. The graph criterion is an incomparability-graph search over finitely many breakpoints $\max(0,-d_{ij},-d_{ji})$. The weighted maximum-flow proof below gives a polynomial-time alternative.

\begin{theorem}[Sharp upper projection]\label{supp:thm:upper}
Let $\pi$ be a permutation of $[m]$ and $C_\pi=C\cap\{x_{\pi(1)}\le\cdots\le x_{\pi(m)}\}$. For each $a$ with $\sum_{k=a}^m w_{\pi(k)}\ge q$, define $h_\pi(a)$ as the smallest $b\ge a$ with $\sum_{k=a}^b w_{\pi(k)}\ge q$. For nonempty $C_\pi$, let $u_\pi$ be the optimum of the linear program
\begin{equation}\label{supp:eq:lp}
 \max_{x\in C_\pi,\ z\ge0}z,\qquad
 z\le x_{\pi(h_\pi(a))}-x_{\pi(a)}\quad\text{for every eligible }a.
\end{equation}
Then $U_q:=\max_{x\in C}R_q(x)=\max_{\pi:C_\pi\ne\varnothing}u_\pi$. Moreover,
\begin{equation}\label{supp:eq:cheap}
 U_q\le \widetilde U_q:=\min_{S\in\mathcal S_q}\max_{i,j\in S}d_{ij}.
\end{equation}
\end{theorem}
The upper bound in (\ref{supp:eq:cheap}) can be computed by a second weighted-clique search, now joining pairs with both directed distances at most the candidate tolerance. The exact upper projection may require a factorial number of programs; it is intended for modest analytical universes. The conservative alternative is valid at any size, although clique optimization can itself be difficult. Neither method assumes that the analyses are independent. Unlike the lower projection, this upper clique problem does not in general have the partial-order structure. For $q=1$, the two upper endpoints coincide with $\max_{i,j}d_{ij}$ (see the argument below). For intermediate weights, they can differ: with three equally weighted points and all pairwise differences bounded by one, the sharp upper endpoint at $q=2/3$ is $1/2$, whereas the conservative endpoint is one. This is the difference between maximizing a minimum and minimizing separate maxima.

\begin{theorem}[Simultaneous validity and inversion]\label{supp:thm:coverage}
Under (\ref{supp:eq:event}), with probability at least $1-\alpha$, simultaneously for all $0<q\le1$ and $t\ge0$,
\[
 L_q\le R_q(\theta)\le U_q,\qquad
 \sup\{q:U_q\le t\}\le A_t(\theta)\le\sup\{q:L_q\le t\}.
\]
The supremum is over $q\in(0,1]$, with the empty supremum defined as zero. Replacing $U_q$ by $\widetilde U_q$ preserves validity.
\end{theorem}
Thus a tolerance $t$ certifies existence of a coherent core of weight $q$ when $U_q\le t$, rules out such a core when $L_q>t$, and otherwise yields an unresolved conclusion. These decisions are jointly protected, even when $q$ and $t$ are selected from the reported profile. They are existential statements; identifying a particular core requires that its own maximum upper contrast bound does not exceed $t$.

\subsection{Calibration and contraction}
Let $\widehat\theta$ denote a joint estimator. If $\widehat\theta-\theta\sim N_m(0,V)$ with known covariance $V$, let $s_{ij}^2=V_{ii}+V_{jj}-2V_{ij}$ and let $c$ be a $(1-\alpha)$ quantile of $\max_{i<j:s_{ij}>0}|G_j-G_i|/s_{ij}$ for $G\sim N_m(0,V)$. If there are no positive variances, take $c=0$. Setting $b_{ij}=\widehat\theta_j-\widehat\theta_i+cs_{ij}$ yields (\ref{supp:eq:event}); zero-variance contrasts are exact. A conservative analytic choice for $m\ge2$ is $c=\Phi^{-1}(1-\alpha/[m(m-1)])$, with $\Phi$ the standard normal distribution function. For estimated covariance and fixed $m$, an analogous plug-in calibration is asymptotically valid if a joint centered Gaussian limit holds, contrast variances are strictly positive, covariance estimates are consistent on the same scaling, and the limiting maximum has a continuous distribution. Resampling must operate on original observational units, not on specifications; validity of the joint approximation must be justified for the particular pipelines.

A distribution-free finite-sample case is $\theta_i=\E f_i(Z)$, $\widehat\theta_i=n^{-1}\sum_{k=1}^n f_i(Z_k)$, for independent identically distributed observations $Z_1,\ldots,Z_n$ and fixed measurable functions $f_i$. Suppose $f_j(Z)-f_i(Z)$ lies almost surely in a known interval of length $B_{ij}=B_{ji}$. By \citet{hoeffding1963} and the union bound, use
\[
 b_{ij}=\widehat\theta_j-\widehat\theta_i+B_{ij}\sqrt{\frac{\log(m(m-1)/\alpha)}{2n}}.
\]
This exploits paired contrast ranges rather than bounding the marginal scores separately. The two-sample extension is derived in Section S2. It does not apply automatically to fitted nonlinear pipelines.

\begin{proposition}[Uniform contraction]\label{supp:prop:rate}
For bounds of the form $b_{ij}=\widehat\theta_j-\widehat\theta_i+e_{ij}$, where $e_{ij}=e_{ji}\ge0$, let $e_*=\max_{i\ne j}e_{ij}$. Then for every $q$, $\widehat R_q-e_*\le L_q\le\widehat R_q\le U_q\le\widetilde U_q\le\widehat R_q+e_*$, where $\widehat R_q=R_q(\widehat\theta)$. Hence the maximum width of either confidence band is at most $2e_*$. In the bounded-score case this is $O(B_*\sqrt{\log(m/\alpha)/n})$, where $B_*=\max B_{ij}$.
\end{proposition}
The rate is a deterministic bound on bandwidth, not a claim of minimax optimality. On the joint coverage event, the target is bracketed even when the minimizing subset changes abruptly, or population effects tie.

\subsection{Proofs of the projection and contraction results}
\begin{proof}[of Theorem~\ref{supp:thm:lower}]
The shortest-path closure gives $x_j-x_i\le d_{ij}$ for every $x\in C$, by summing along paths. Conversely, these bounds imply the original edge inequalities because $d_{ij}\le b_{ij}$. Fix nonempty $S$ and $r\ge0$. If some $d_{ij}<-r$ for $i,j\in S$, then $x_i-x_j>r$ for all feasible $x$, excluding a range at most $r$. Conversely suppose $d_{ij}\ge-r$ for every $i,j\in S$. Add directed edges of length $r$ between all distinct vertices of $S$. Any cycle without new edges is nonnegative. A cycle with new edges can be decomposed into $k\ge1$ new edges and $k$ intervening old paths (possibly empty) whose endpoints lie in $S$. Each old path has length at least $-r$, so the total length is at least zero. Thus the augmented graph has no negative cycle. Add a source with zero-length edges to all vertices; shortest-path distances from this source yield a feasible potential by the triangle inequality. Subtract its first coordinate to anchor it. The new edges ensure the range on $S$ is at most $r$. Taking $r=\max_{i,j\in S}(-d_{ij})\ge0$ gives the attained minimum range for $S$. The two finite/compact minimizations commute, proving (\ref{supp:eq:lower}); the clique criterion follows by pairwise equivalence.
\end{proof}
\begin{proof}[of Theorem~\ref{supp:thm:upper}]
On $C_\pi$, any eligible subset with smallest ordered index $a$ and largest $b$ has weight at most that of the contiguous block $a,\ldots,b$. Consequently $b\ge h_\pi(a)$, and the subset range is at least the corresponding minimal-block range. Conversely, each minimal block is eligible. Thus $R_q(x)$ equals the minimum of the linear block ranges in (\ref{supp:eq:lp}). Its maximum on $C_\pi$ is the displayed linear program, which attains its optimum by compactness. The permutation cells cover $C$, including ties, proving the identity. For each fixed $S$, $R_q(x)\le\max_{i,j\in S}(x_j-x_i)\le\max_{i,j\in S}d_{ij}$. Minimizing the last expression and maximizing the first proves (\ref{supp:eq:cheap}).
\end{proof}
\begin{proof}[of Theorem~\ref{supp:thm:coverage}]
On (\ref{supp:eq:event}), $\theta-\theta_1\mathbf1\in C$, where $\mathbf1$ is the vector of ones. Translation invariance and the definition of extrema give the profile bounds for all $q$ on this single event. By (\ref{supp:eq:mass}), $R_q\le t$ is equivalent to $q\le A_t$. Therefore $U_q\le t$ implies $q\le A_t$, and every $q\le A_t$ satisfies $L_q\le t$. Taking suprema establishes the inversion. The conservative replacement follows from Theorem~\ref{supp:thm:upper}.
\end{proof}
\begin{proof}[of Proposition~\ref{supp:prop:rate}]
The centered estimate belongs to $C$. For any $x\in C$, the paired constraints give $|(x_j-x_i)-(\widehat\theta_j-\widehat\theta_i)|\le e_*$. Each subset range is the maximum of its ordered contrasts, including zero diagonal contrasts. Maxima and minima of finite collections are nonexpansive in the supremum norm, so $|R_q(x)-\widehat R_q|\le e_*$. Also $d_{ij}\le b_{ij}$ for distinct vertices. Choosing a subset that minimizes the empirical range shows $\widetilde U_q\le\widehat R_q+e_*$. The remaining inequalities follow from inclusion of the centered estimate and Theorem~\ref{supp:thm:upper}. The rate follows by substitution.
\end{proof}
\noindent\textit{Full-weight identity and endpoint attainment.} For $q=1$, the only eligible set is $[m]$, so $R_1(x)=\max_{i,j}(x_j-x_i)$. Shortest-path closure gives $\max_{x\in C}(x_j-x_i)\le d_{ij}$. Equality holds: set $y_k=d_{ik}$ and $x_k=y_k-y_1$. The triangle inequality $d_{i\ell}\le d_{ik}+b_{k\ell}$ ensures $x\in C$; $x_j-x_i=d_{ij}$ because $d_{ii}=0$. Maximization over the finite set of pairs and over $C$ commutes. Thus $U_1=\widetilde U_1=\max_{i,j}d_{ij}$. For $q\le\max_i w_i$, a singleton is eligible and all endpoints are identically zero. This identity explains the degenerate portion of the heavily weighted earnings profile.

\subsection{The maximum-flow identity and its proof}\label{supp:sec:flowproof}
The lower projection theorem identifies the compatibility graph as the incomparability graph of a partial order. For $t\ge0$ write $i\prec_t j$ when $d_{ji}<-t$. As $d_{ii}=0$, the relation is irreflexive. If $i\prec_t j$ and $j\prec_t k$, then $d_{ki}\le d_{kj}+d_{ji}<-2t\le-t$, proving transitivity. Thus a set $S$ with $\max_{i,j\in S}(-d_{ij})\le t$ is exactly an antichain. By the lower projection theorem, this is equivalent to feasibility of $\diam_S(x)\le t$ somewhere in $C$, where $\diam_S(x)=\max_{i,j\in S}(x_j-x_i)$.

Consider the network with source $s$, sink $v$, left and right copies of each vertex, capacities $w_i$ on $s\to i_L$ and $i_R\to v$, and capacity 2 on $i_L\to j_R$ whenever $i\prec_t j$. Let $F_t$ be its maximum flow. Denote by $\gamma_{ij}$ the flow along each middle edge and by $o_i=\sum_j\gamma_{ij}$ and $a_i=\sum_j\gamma_{ji}$ its total outgoing and incoming flow at the original vertex. Both $o_i$ and $a_i$ are at most $w_i$. The directed order graph is acyclic. Add an auxiliary source sending $w_i-a_i$ units into vertex $i$ and an auxiliary sink receiving $w_i-o_i$ units from it. At every original vertex, the resulting flow is conserved and has throughput $w_i$. Decompose this acyclic flow into source-to-sink paths with nonnegative weights. The total path weight equals $\sum_i(w_i-a_i)=1-F_t$. Each path passes through a chain of comparable vertices, and any antichain meets each chain at most once. Consequently the weight of every antichain is at most $1-F_t$.

Conversely, a minimum cut has capacity $F_t\le1$, so it cuts no middle edge of capacity 2. Let $P$ be the vertices whose left copies lie on the sink side, and $Q$ those whose right copies lie on the source side. Its capacity is $w(P)+w(Q)=F_t$ by the max-flow/min-cut theorem \citep{ford1956}. The set $A=[m]\setminus(P\cup Q)$ is an antichain: if $i,j\in A$ and $i\prec_t j$, a middle edge would cross the cut. Its weight satisfies $w(A)\ge1-w(P)-w(Q)=1-F_t$. This proves equality for arbitrary positive real weights, and gives a constructive optimal antichain. The argument is a weighted form of the classical chain--antichain relationship \citep{dilworth1950}. Together with the feasibility statement, this gives $\max_{x\in C}A_t(x)=1-F_t$ and the formula for $L_q$. Since $d_{ij}$ are fixed once the region is observed, the threshold search has at most $1+\binom m2$ candidate values, with strict inequalities used in defining the order so that boundary ties are retained. Binary search is valid by monotonicity. Standard polynomial-time max-flow algorithms give a polynomial bound in the number of vertices and edges (and polynomial bit complexity for rational inputs). No such assertion is made for the upper projection. The audit below uses integer capacities and distances to check the boundary convention exactly.

\subsection{A family with a large upper-projection gap}
Let the weights be equal, and let $C$ be the anchored region with $|x_j-x_i|\le D$ for all pairs, for some $D>0$. Then $d_{ij}=D$ for distinct vertices and $d_{ii}=0$. For $q=k/m$, $2\le k\le m$, the separate-subset bound $\widetilde U_q$ equals $D$. Let $p=\lfloor(m-1)/(k-1)\rfloor$. In sorted order, the $p$ blocks joining ranks $1+\ell(k-1)$ and $1+(\ell+1)(k-1)$, $\ell=0,\ldots,p-1$, have ranges each at least $R_q(x)$ and total range at most $D$. Thus $U_q\le D/p$. The ordered configuration $x_i=D\lfloor(i-1)/(k-1)\rfloor/p$ lies in $C$ and every contiguous $k$-block has range exactly $D/p$, giving equality. This proves the formula in the article. The ratio $p$ can grow without bound (take $k=2$ and increase $m$), although both bounds remain valid confidence endpoints. This is a deterministic example, not a claim about the typical size of the gap in a sampling experiment.

\subsection{Directional limits at competing subsets}\label{supp:sec:direction}
The projection method does not require the target to be differentiable. The following local expansion makes explicit the problem with direct smooth approximations. For each $S\in\mathcal S_q$ define
\[
 I^+_S(\theta)=\{i\in S:\theta_i=\max_{j\in S}\theta_j\},\quad
 I^-_S(\theta)=\{i\in S:\theta_i=\min_{j\in S}\theta_j\},
\]
and let $\mathcal M_q(\theta)=\{S\in\mathcal S_q:\diam_S(\theta)=R_q(\theta)\}$. For fixed $m$ the directional derivative in direction $h\in\R^m$ is
\begin{equation}\label{supp:eq:directional}
 R'_{q,\theta}(h)=\min_{S\in\mathcal M_q(\theta)}
 \{\max_{i\in I^+_S(\theta)}h_i-\min_{j\in I^-_S(\theta)}h_j\}.
\end{equation}
To prove this, take any $u_n\downarrow0$ and $h_n\to h$. Since each finite set of nonmaximizers has a strictly positive gap from the maximum (unless empty), the maximum of $\theta_i+u_n h_{ni}$ over a fixed $S$ is attained within $I^+_S(\theta)$ for sufficiently large $n$; the same holds for the minimum. Hence for each $S$ the scaled range change tends to the corresponding difference of active extremes. Nonoptimal subsets have strictly positive range gaps, and all the range perturbations are bounded by $2u_n\|h_n\|_\infty$. They therefore cannot attain the outer minimum as $n$ grows large. The finite minimum over the active subsets gives (\ref{supp:eq:directional}). The same argument works jointly over the finite set of positive attainable subset weights. This establishes Hadamard directional differentiability in finite dimension. If $\sqrt n(\widehat\theta-\theta)\rightsquigarrow G$ for a Gaussian vector $G$, the directional delta method \citep{fang2019} gives joint convergence of the scaled profile errors to the vector of derivatives in (\ref{supp:eq:directional}). At complete agreement, the derivative is $R_q(h)$ itself; if $q$ exceeds every singleton weight and all pair contrasts of $G$ have positive variance, $R_q(G)>0$ almost surely, so the limit is nonnegative and generally non-Gaussian. If a unique active subset has unique extremes, the derivative is their linear contrast. At other configurations, it can be nonlinear even for Gaussian input. The argument is a fixed-universe limit and does not assert uniform bootstrap validity near ties. The simultaneous contrast approach avoids the need to estimate active sets.

\subsection{Independent check of the flow formulation}
The accompanying script \texttt{verify\_flow.py} constructs 120 feasible, asymmetric integer difference-constraint regions, of dimensions 2--9, with positive integer weights normalized to sum to one. It compares the maximum-flow and exhaustive-subset values at each candidate threshold and reconstructs the lower profile at every attainable weight. The check uses exact integer capacities and verifies the strict-threshold convention in ties. These deterministic-seed algorithm checks are separate from the 24,000 Monte Carlo replications described in Section S3. All 1,144 threshold comparisons and 2,011 profile comparisons agreed exactly. The output file \texttt{audit/flow\_audit.json} records the checks; no additional coverage claims are based on this verification.

\section{Calibration for the two-sample experiment}\label{supp:sec:two}
Let $Y_{ak}$, $k=1,\ldots,n_a$, be independent samples from the distributions in arms $a=0,1$, with independence between arms. With fixed measurable functions $f_i$, write $g_{ij}=f_j-f_i$ and suppose its values lie in an interval of known length $B_{ij}$. The contrast estimator $\widehat\theta_j-\widehat\theta_i$ is the difference of arm-specific sample means of $g_{ij}$. Each summand in arm 1 has range length $B_{ij}/n_1$, and each summand in arm 0 has range length $B_{ij}/n_0$. Hoeffding's inequality \citep{hoeffding1963} and the union bound over $m(m-1)$ ordered pairs give the simultaneous radius
\[
e_{ij}=B_{ij}\sqrt{\frac{1}{2}\left(\frac1{n_1}+\frac1{n_0}\right)\log\frac{m(m-1)}{\alpha}}.
\]
For $Y\ge0$ and finite caps $c_i,c_j$, $B_{ij}=|c_j-c_i|$ follows directly from $\min(Y,c_j)-\min(Y,c_i)$. No known finite range is assumed for an uncapped comparison. For $a<b$ finite, $\min(Y,b)-\min(Y, a)=\int_a^b\mathbf1\{Y>u\}\,du$; integrating the arm expectations proves the survival-contrast identity in the main article. For an infinite upper cap, it extends by integrability of the earnings distributions. This provides a population interpretation of cap sensitivity, not an assumption that capped and uncapped effects coincide.

For Gaussian simulation calibration, with $K=120000$ simulated independent Gaussian maxima, the code selects the order statistic of rank $\lceil(K+1)(1-\alpha)\rceil$. With a continuous maximum and a fresh independent data maximum, the unconditional probability of falling below this order statistic is its rank divided by $K+1$. In the simulations, a single calibration value is reused within each scenario; Monte Carlo coverage frequencies are therefore conditional on that realized calibration, and can differ slightly from nominal even for the joint contrast region. The actual profile event can be strictly larger than the contrast event. For real data, the same Gaussian procedure substitutes $\widehat V$, and the bootstrap draws participants separately within each arm, calculates centered mean contrasts, and standardizes them by the original estimated contrast standard errors. Both are asymptotic approximations under the stated independent two-sample model, not exact randomization tests. The bounded-score method is finite-sample under the specified sampling and range assumptions.

\section{Extended Monte Carlo design and performance}\label{supp:sec:sim}
The Gaussian design uses $\Sigma=D\Omega D$ with $D_{ii}$ evenly spaced from 0.75 to 1.25. For $m=6$, the three mean configurations, covariance structures, and sample sizes are as described in the article. For $m=10$, the coherent core comprises eight equally spaced values from 0 to 0.04 and two values, 0.45 and 0.52; the common dependence structure and both sample sizes are used. Four bounded-score experiments use $m=6$, normalized weights proportional to $(1,2,1,2,1,1)$, the tie or core mean configuration, and $n=100,500$. Specifically,
\[
X_{ki}=\theta_i+2H_k+a_iE_{ki},\qquad a_i=0.25+0.30(i-1)/(m-1),
\]
where all $H_k$ and $E_{ki}$ are independent Rademacher variables (each takes $\pm1$ with probability one-half). The covariance of a single observation is $4\mathbf1\mathbf1^\top+\operatorname{diag}(a_i^2)$ and the paired score contrast has known range length $2(a_i+a_j)$ for $i\ne j$. The marginal box and paired normal methods in this non-Gaussian design are approximate, even when covariance is known.

For each replication, all distinct positive subset weights are evaluated; with integer weights, this is implemented by enumerating achievable sums and dividing by their total. Write $\mathcal Q$ for this set. A replication is simultaneously covered when all $q\in\mathcal Q$ are covered. The reported mean width averages $U_q-L_q$ (using the conservative upper endpoint in the main experiment) over $\mathcal Q$, then over replications. The raw files also record bias, pointwise coverage, and incorrect certification or rejection at tolerance 0.15; the corresponding replication-level indicator is one if any attainable weight produces an erroneous decision. The selected pointwise comparator uses the empirically selected endpoints and a normal interval with the known paired standard error, truncated below at zero. It is a diagnostic, not a procedure with a simultaneous guarantee. The reported marginal box uses Bonferroni marginal normal intervals, and working independence replaces the true covariance with a diagonal covariance when constructing contrast errors. In each case, the pairwise bounds are projected as described above, except for the selected pointwise diagnostic.

\begin{table}[!htbp]\centering\scriptsize
\caption{All Gaussian scenarios. Entries are simultaneous coverage / mean profile width, each based on 1,000 replications. M: paired maxT; B: paired Bonferroni; E: estimated-covariance Bonferroni; X: marginal box; I: working independence; P: selected pointwise. Width includes structurally zero breakpoints.}\label{supp:tab:fullgauss}
\resizebox{\linewidth}{!}{\begin{tabular}{lllrrcccccc}\toprule ID&Shape&Dependence&$n$&$m$&M&B&E&X&I&P\\\midrule
S01 & tie & independent & 100 & 6 & 0.950 / 0.430 & 0.960 / 0.442 & 0.954 / 0.441 & 0.998 / 0.533 & 0.960 / 0.442 & 0.667 / 0.323 \\
S02 & tie & independent & 500 & 6 & 0.957 / 0.192 & 0.969 / 0.197 & 0.966 / 0.197 & 1.000 / 0.238 & 0.969 / 0.197 & 0.688 / 0.145 \\
S03 & tie & common & 100 & 6 & 0.951 / 0.155 & 0.965 / 0.159 & 0.949 / 0.159 & 1.000 / 0.471 & 1.000 / 0.379 & 0.675 / 0.117 \\
S04 & tie & common & 500 & 6 & 0.951 / 0.069 & 0.967 / 0.071 & 0.968 / 0.071 & 1.000 / 0.211 & 1.000 / 0.170 & 0.691 / 0.052 \\
S05 & tie & signed & 100 & 6 & 0.954 / 0.419 & 0.977 / 0.441 & 0.964 / 0.441 & 0.990 / 0.525 & 0.944 / 0.435 & 0.825 / 0.332 \\
S06 & tie & signed & 500 & 6 & 0.960 / 0.188 & 0.979 / 0.198 & 0.982 / 0.198 & 0.987 / 0.236 & 0.939 / 0.195 & 0.812 / 0.149 \\
S07 & core & independent & 100 & 6 & 0.994 / 0.510 & 0.994 / 0.525 & 0.993 / 0.524 & 0.999 / 0.637 & 0.994 / 0.525 & 0.851 / 0.369 \\
S08 & core & independent & 500 & 6 & 0.996 / 0.231 & 0.997 / 0.237 & 0.997 / 0.237 & 0.999 / 0.292 & 0.997 / 0.237 & 0.885 / 0.166 \\
S09 & core & common & 100 & 6 & 0.988 / 0.193 & 0.993 / 0.199 & 0.989 / 0.198 & 1.000 / 0.603 & 1.000 / 0.494 & 0.902 / 0.138 \\
S10 & core & common & 500 & 6 & 0.996 / 0.091 & 0.997 / 0.094 & 0.997 / 0.094 & 1.000 / 0.275 & 1.000 / 0.222 & 0.905 / 0.066 \\
S11 & core & signed & 100 & 6 & 0.992 / 0.479 & 0.998 / 0.507 & 0.995 / 0.506 & 0.998 / 0.625 & 0.975 / 0.515 & 0.911 / 0.364 \\
S12 & core & signed & 500 & 6 & 0.992 / 0.220 & 0.995 / 0.233 & 0.995 / 0.234 & 0.996 / 0.291 & 0.972 / 0.236 & 0.870 / 0.167 \\
S13 & gradient & independent & 100 & 6 & 1.000 / 0.517 & 1.000 / 0.530 & 1.000 / 0.530 & 1.000 / 0.631 & 1.000 / 0.530 & 0.966 / 0.384 \\
S14 & gradient & independent & 500 & 6 & 0.992 / 0.268 & 0.994 / 0.277 & 0.993 / 0.277 & 1.000 / 0.337 & 0.994 / 0.277 & 0.861 / 0.192 \\
S15 & gradient & common & 100 & 6 & 0.984 / 0.241 & 0.990 / 0.250 & 0.985 / 0.249 & 1.000 / 0.638 & 1.000 / 0.545 & 0.800 / 0.171 \\
S16 & gradient & common & 500 & 6 & 0.968 / 0.115 & 0.976 / 0.120 & 0.975 / 0.120 & 1.000 / 0.349 & 1.000 / 0.289 & 0.756 / 0.080 \\
S17 & gradient & signed & 100 & 6 & 0.995 / 0.520 & 0.998 / 0.548 & 0.997 / 0.547 & 0.996 / 0.626 & 0.982 / 0.526 & 0.761 / 0.411 \\
S18 & gradient & signed & 500 & 6 & 0.983 / 0.289 & 0.992 / 0.306 & 0.992 / 0.306 & 0.994 / 0.340 & 0.966 / 0.280 & 0.747 / 0.217 \\
S23 & core & common & 100 & 10 & 0.995 / 0.204 & 0.998 / 0.210 & 0.996 / 0.210 & 1.000 / 0.622 & 1.000 / 0.527 & 0.837 / 0.137 \\
S24 & core & common & 500 & 10 & 0.997 / 0.096 & 0.998 / 0.099 & 0.998 / 0.099 & 1.000 / 0.283 & 1.000 / 0.237 & 0.889 / 0.065 \\

\bottomrule\end{tabular}}
\end{table}
\begin{table}[!htbp]\centering\scriptsize
\caption{All bounded-score scenarios, using the same summary as Table~\ref{supp:tab:fullgauss}. H denotes paired Hoeffding; other abbreviations are unchanged.}\label{supp:tab:fullbounded}
\resizebox{\linewidth}{!}{\begin{tabular}{lllrrcccccc}\toprule ID&Shape&Dependence&$n$&$m$&H&B&E&X&I&P\\\midrule
S19 & tie & common & 100 & 6 & 1.000 / 0.247 & 0.975 / 0.160 & 0.963 / 0.160 & 1.000 / 0.843 & 1.000 / 0.671 & 0.658 / 0.117 \\
S20 & core & common & 100 & 6 & 1.000 / 0.296 & 0.997 / 0.185 & 0.994 / 0.185 & 1.000 / 0.953 & 1.000 / 0.782 & 0.886 / 0.130 \\
S21 & tie & common & 500 & 6 & 1.000 / 0.111 & 0.960 / 0.072 & 0.956 / 0.072 & 1.000 / 0.377 & 1.000 / 0.300 & 0.657 / 0.053 \\
S22 & core & common & 500 & 6 & 1.000 / 0.133 & 0.991 / 0.085 & 0.993 / 0.085 & 1.000 / 0.491 & 1.000 / 0.392 & 0.879 / 0.060 \\

\bottomrule\end{tabular}}
\end{table}
Table~\ref{supp:tab:fullgauss} shows that the maxT profile coverages in the two ten-specification experiments were 0.995 and 0.997, with mean widths 0.204 and 0.096 at sample sizes 100 and 500. Across the 18 six-specification Gaussian cases, the mean widths under maxT fell from 0.196 to 0.092 under common dependence when sample size increased from 100 to 500 (averaging over shapes). The corresponding widths under independent coordinates were 0.486 and 0.230. This is compatible with root-sample-size contraction but is not an empirical proof of a rate. Table~\ref{supp:tab:fullbounded} shows the distinction between finite-sample bounded-score protection and approximate paired normal calibration. At both sample sizes, the Hoeffding construction retains the benefit of canceling the common score term, yet remains conservative.

\begin{figure}[p]\centering\includegraphics[width=\linewidth]{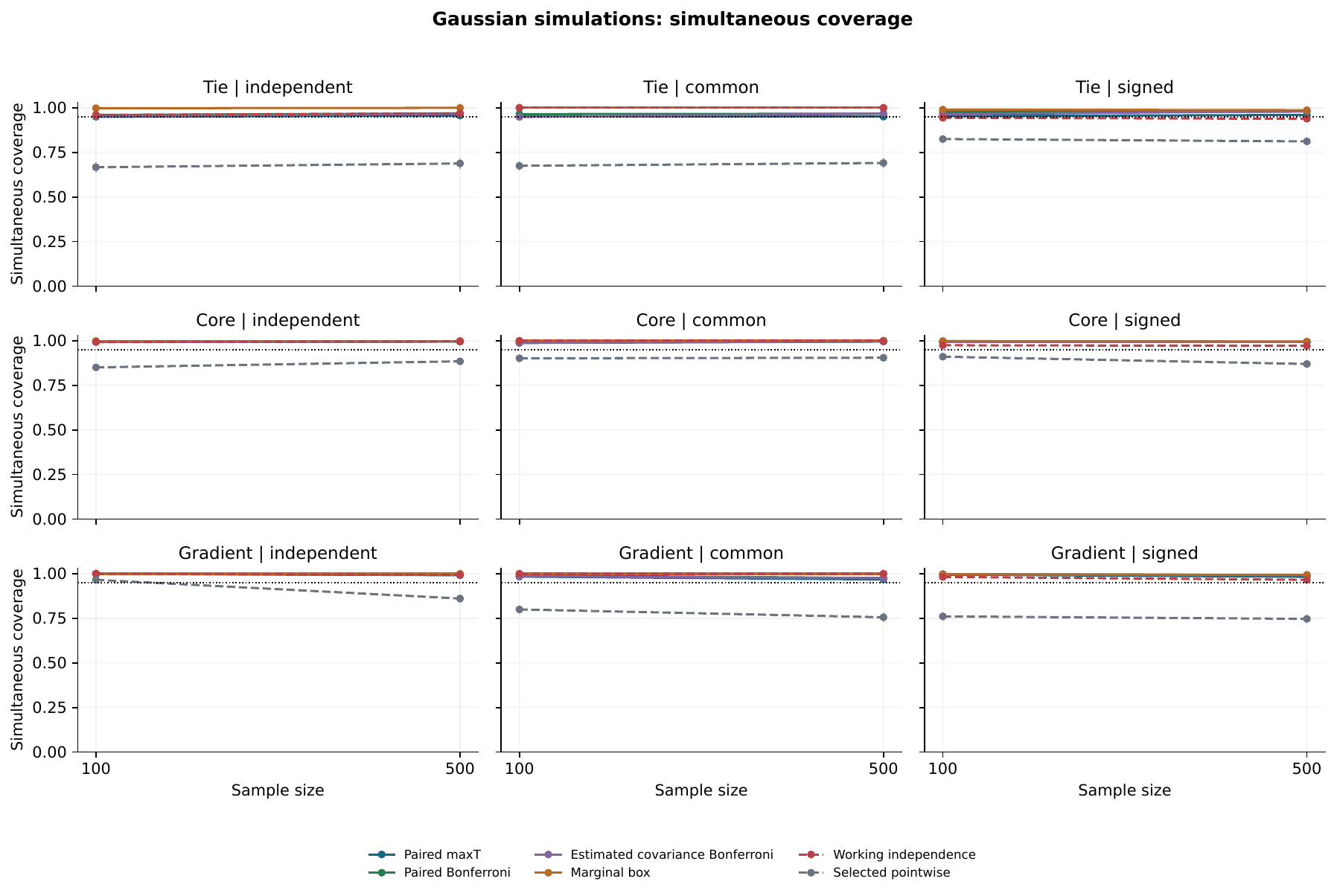}\caption{Scenario-specific simultaneous coverage for the six-specification Gaussian design. Error bars are 95\% Wilson intervals for Monte Carlo proportions; the reference line is 0.95.}\label{supp:fig:fullcover}\end{figure}
\begin{figure}[!htbp]\centering\includegraphics[width=\linewidth]{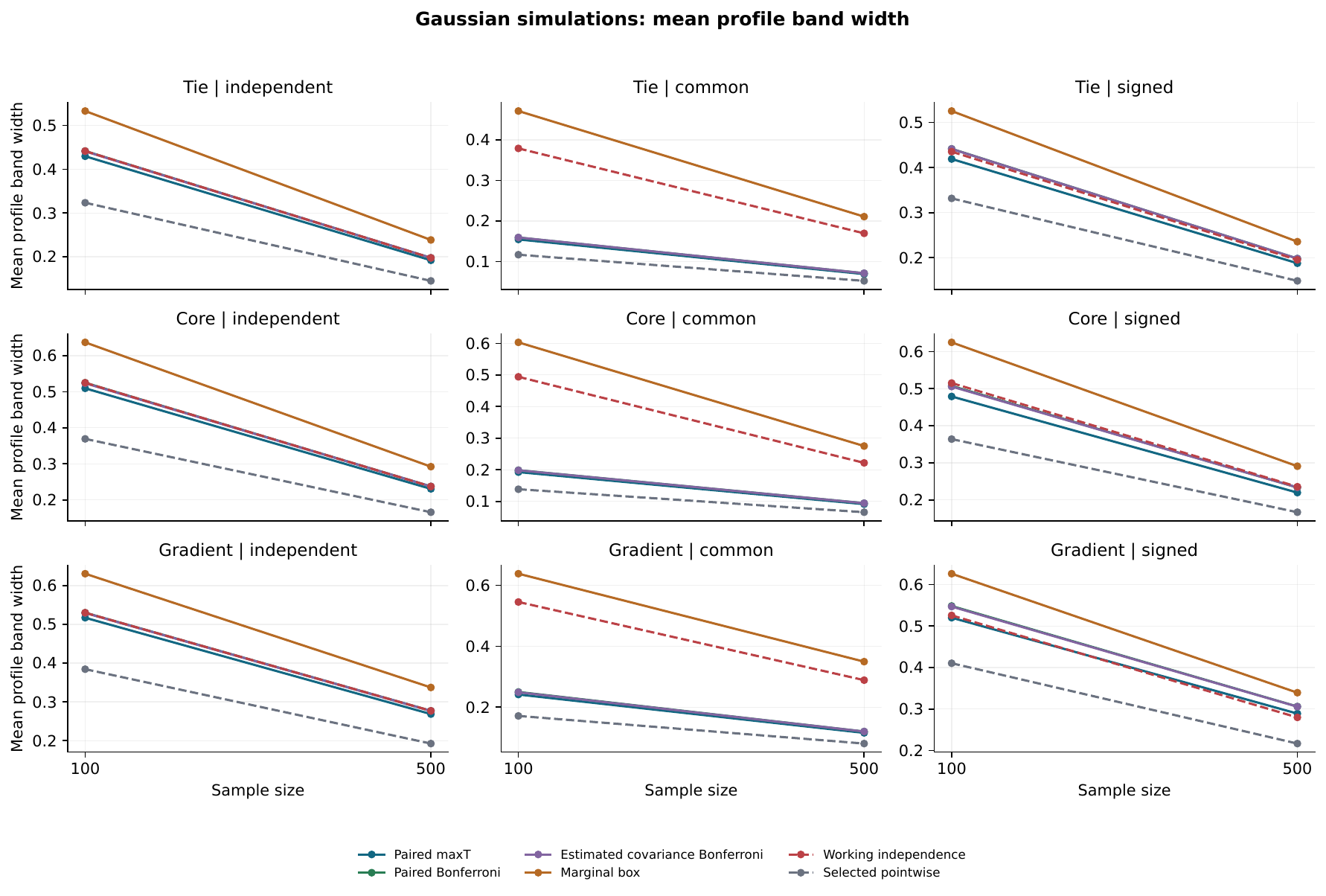}\caption{Scenario-specific average profile widths for the same design as Figure~\ref{supp:fig:fullcover}. A short band is not evidence of good calibration when simultaneous coverage is poor.}\label{supp:fig:fullwidth}\end{figure}
\begin{figure}[!htbp]\centering
\begin{subfigure}{\linewidth}\includegraphics[width=\linewidth]{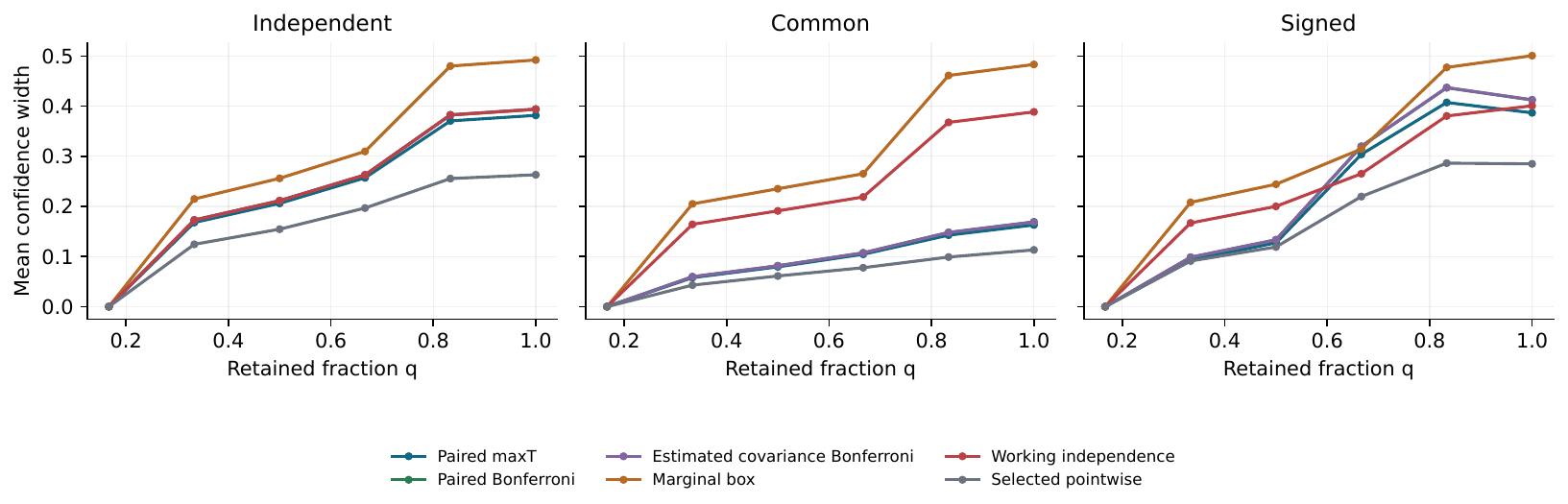}\caption{Widths at individual retained-weight breakpoints in the Gaussian core configuration, $n=500$.}\label{supp:fig:qwidth}\end{subfigure}\par\medskip
\begin{subfigure}{\linewidth}\includegraphics[width=\linewidth]{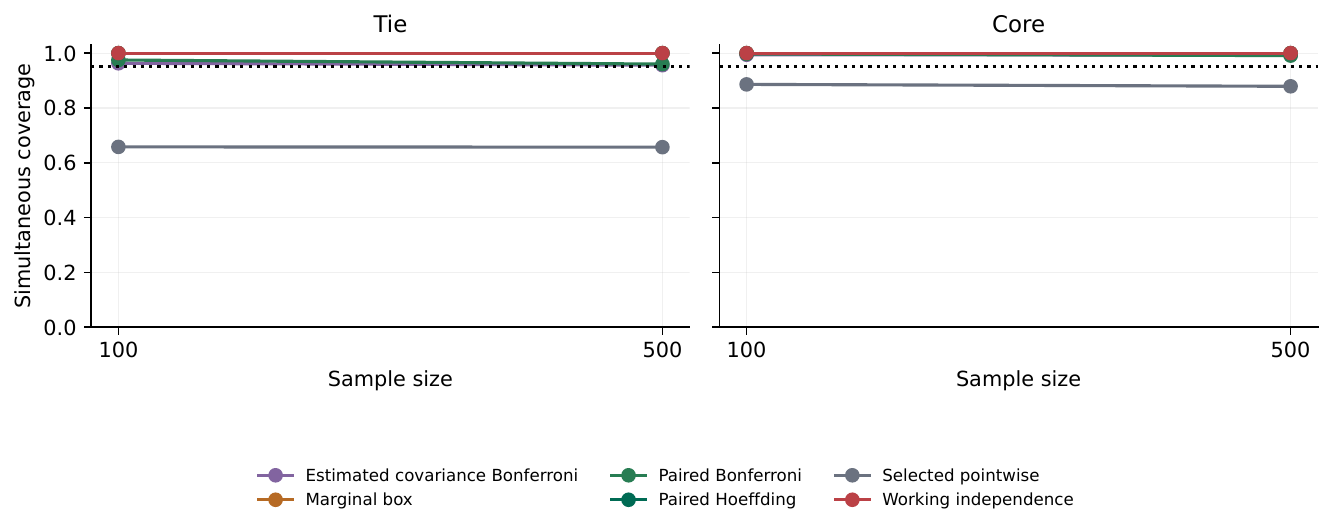}\caption{Simultaneous coverage under bounded scores with unequal weights.}\label{supp:fig:bounded}\end{subfigure}
\caption{Weight-specific precision and bounded-score coverage.}\label{supp:fig:additional_sim}
\end{figure}
Figures~\ref{supp:fig:fullcover}--\ref{supp:fig:additional_sim} resolve the aggregated main-text comparison by configuration, sample size, and retained weight. In particular, the structurally zero singleton breakpoints must not be interpreted as a successful noisy estimate. The core case has comparatively high coverage even when some individual contrasts miss, illustrating the difference between joint-region and profile coverage. Full pointwise statistics, Wilson limits, and simultaneous diagnostics are available in the accompanying audit CSV files (see Section~\ref{supp:sec:repro}).

\clearpage
\section{Numerical verification and exact-projection study}\label{supp:sec:audit}
We checked 108 asymmetric feasible regions: three dimensions ($m=4,5,6$), three mean configurations, and 12 replications per combination. A mean vector was perturbed by independent Gaussian noise of standard deviation 0.04, and independent directed slack terms were sampled uniformly between 0.025 and 0.25 for off-diagonal constraints. Alternating replications used equal or unequal weights (the latter proportional to $(1,2,1,\ldots,1)$). Two attainable weights closest to 0.6 and 0.8 were checked in each region. The independent lower oracle enumerates eligible subsets, minimizes an auxiliary range variable via direct linear programming, and then selects the smallest optimal value. The exact upper bound enumerates feasible orderings and checks the feasibility of the returned optimal configurations. The lower endpoints agreed within $1.12\times10^{-16}$ across all 216 evaluations. Upper gaps were positive in 96 evaluations (44.4\%), reaching 0.078; this is numerical evidence supporting the formulas, not a substitute for their proofs.

\begin{figure}[!htbp]\centering\includegraphics[width=\linewidth]{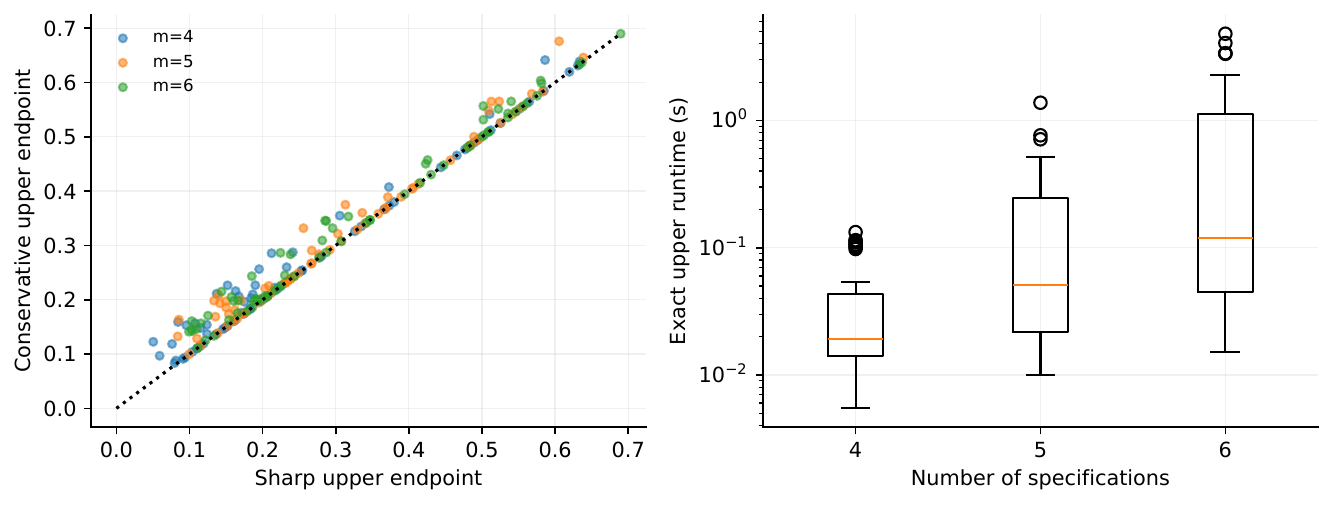}\caption{Numerical audit. Left: conservative versus sharp upper endpoints over 216 evaluations; points above the identity line reflect nonzero projection gaps. Right: elapsed time for the exact upper calculation at two weights per region, by dimension, on the recorded Colab runtime.}\label{supp:fig:lpaudit}\end{figure}
Figure~\ref{supp:fig:lpaudit} documents the computational trade-off: median exact-search elapsed times per region were approximately 0.019, 0.051 and 0.120 seconds for four, five and six specifications, respectively, compared with roughly 0.00026--0.00032 seconds for the lower and conservative upper calculations (all weights). Timings are machine-dependent and do not establish algorithmic complexity. The analytic three-point example returned exactly one-half, and the negative-cycle, translation-invariance, and weight-splitting checks passed. These tests exercise the geometry as well as calibration-independent feasibility.

\begin{figure}[!htbp]\centering\includegraphics[width=\linewidth]{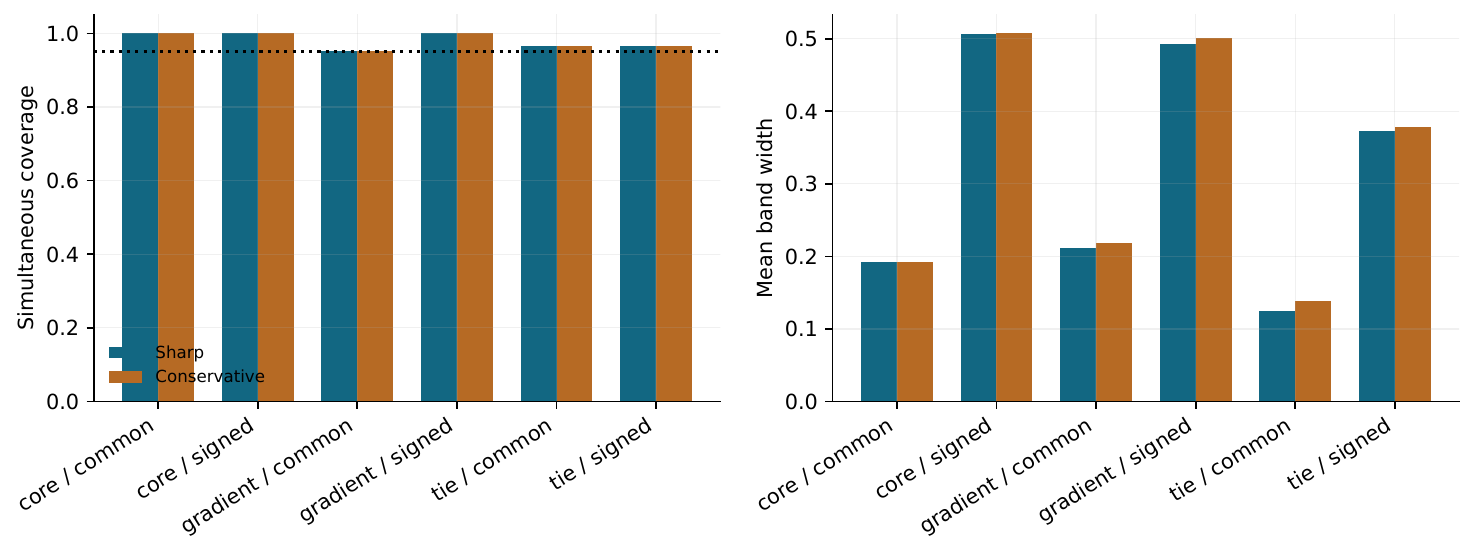}\caption{Four-specification exact-band simulation. Left: simultaneous coverage; right: mean width averaged over all attainable weights. The two bands have the same lower endpoint.}\label{supp:fig:exact}\end{figure}

The separate exact-band experiment uses $m=4$, equal weights, $n=100$, known Gaussian covariance with common or signed dependence, three mean configurations, and 150 replications per scenario (900 in total). The Bonferroni critical value is used for both bands. Table~\ref{supp:tab:exact} and Figure~\ref{supp:fig:exact} report the comparison. At half weight, 48.1\% of replicates exhibit a strictly positive gap (tolerance $10^{-8}$), whereas at full weight the gap is identically zero by the full-weight identity. The maximum gap over all weights in this experiment is 0.212. Equality of observed coverage across the two procedures is compatible with strict improvement in width.
\begin{table}[!htbp]\centering\small\caption{Exact versus conservative bands: coverage of the full profile and mean width over its attainable weights. Each row uses 150 replications.}\label{supp:tab:exact}
\begin{tabular}{llrrrr}\toprule&&\multicolumn{2}{c}{Coverage}&\multicolumn{2}{c}{Mean width}\\\cmidrule(lr){3-4}\cmidrule(lr){5-6}Shape&Dependence&Sharp&Conservative&Sharp&Conservative\\\midrule
Core&Common&1.000&1.000&0.193&0.193\\
Core&Signed&1.000&1.000&0.507&0.507\\
Gradient&Common&0.953&0.953&0.211&0.218\\
Gradient&Signed&1.000&1.000&0.493&0.501\\
Tie&Common&0.967&0.967&0.124&0.139\\
Tie&Signed&0.967&0.967&0.372&0.379\\\bottomrule\end{tabular}\end{table}

\clearpage
\section{Further analysis of experimental earnings}\label{supp:sec:earnings}
The National Supported Work subsample \citep{lalonde1986,dehejia1999,dehejia2002} was retrieved from the Rdatasets distribution \citep{nswmixtape}, with the original archive documented by \citet{nswarchive}. The validated sample contains 185 treated and 260 control observations; the uncapped difference of 1794.342 historical dollars is reproduced. The input SHA-256 checksum is recorded in \texttt{audit/data\_provenance.json}. All monetary values below are in thousands of historical US dollars. The covariances use the usual sample covariance, with the denominator $n_a-1$, separately for each arm.

\begin{figure}[!htbp]\centering
\begin{subfigure}[t]{.52\linewidth}\centering\includegraphics[width=\linewidth]{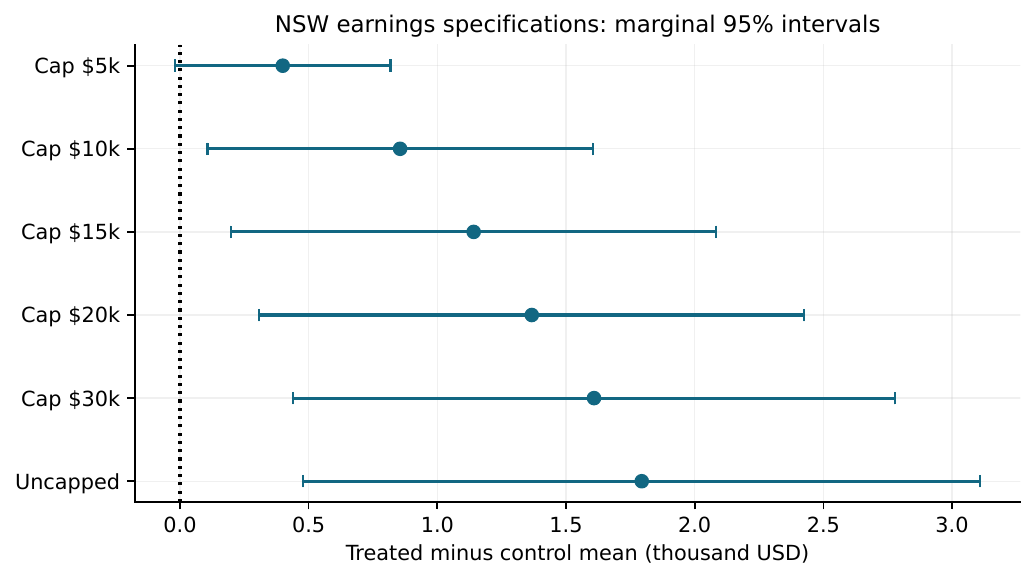}\caption{Estimates and marginal normal intervals.}\label{supp:fig:nswspec}\end{subfigure}\hfill
\begin{subfigure}[t]{.44\linewidth}\centering\includegraphics[width=\linewidth]{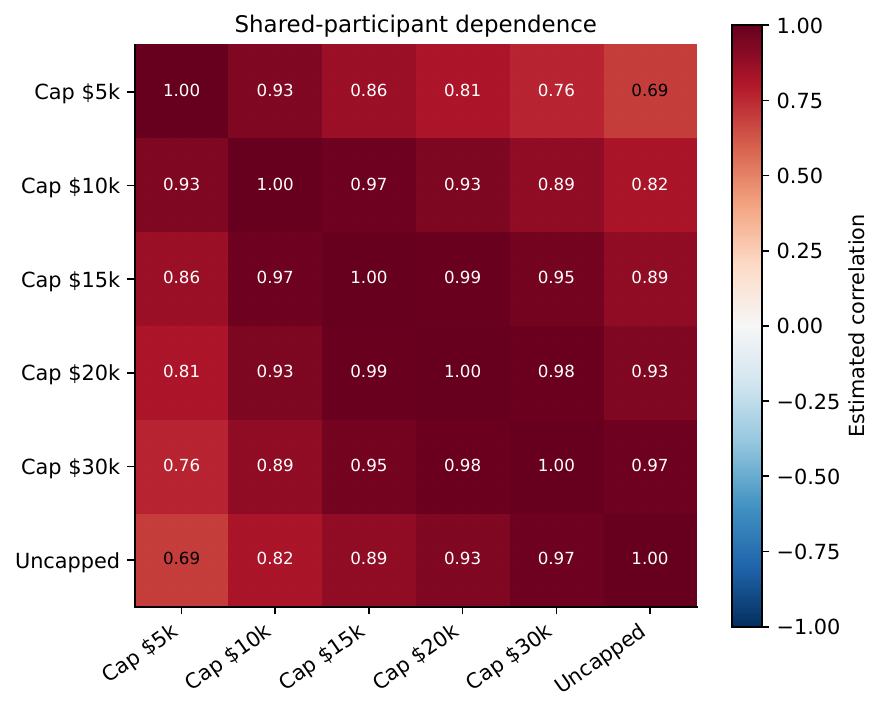}\caption{Correlation of estimated effects.}\label{supp:fig:nswcorr}\end{subfigure}
\caption{The earnings outcome definitions and their shared-participant dependence. The intervals in panel (a) are marginal, not simultaneous.}\label{supp:fig:nswdeps}\end{figure}
Figure~\ref{supp:fig:nswdeps} shows why treating specifications as separate studies is inappropriate. The 15- and 20-thousand-dollar cap estimates have a paired standard error of about 0.099, compared with 0.723 under the working independence assumption. The proportion of treated participants above the 5-thousand-dollar cap is 45.4\%, versus 37.7\% for controls; at the 30-thousand-dollar cap, the proportions are 1.6\% and 0.4\%, respectively. This makes the estimand changes transparent. The joint confidence region nevertheless permits a wide range of population-level tail contributions. The sample covariance and all 15 contrast standard errors are in the accompanying audit tables.

\begin{figure}[!htbp]\centering\includegraphics[width=\linewidth]{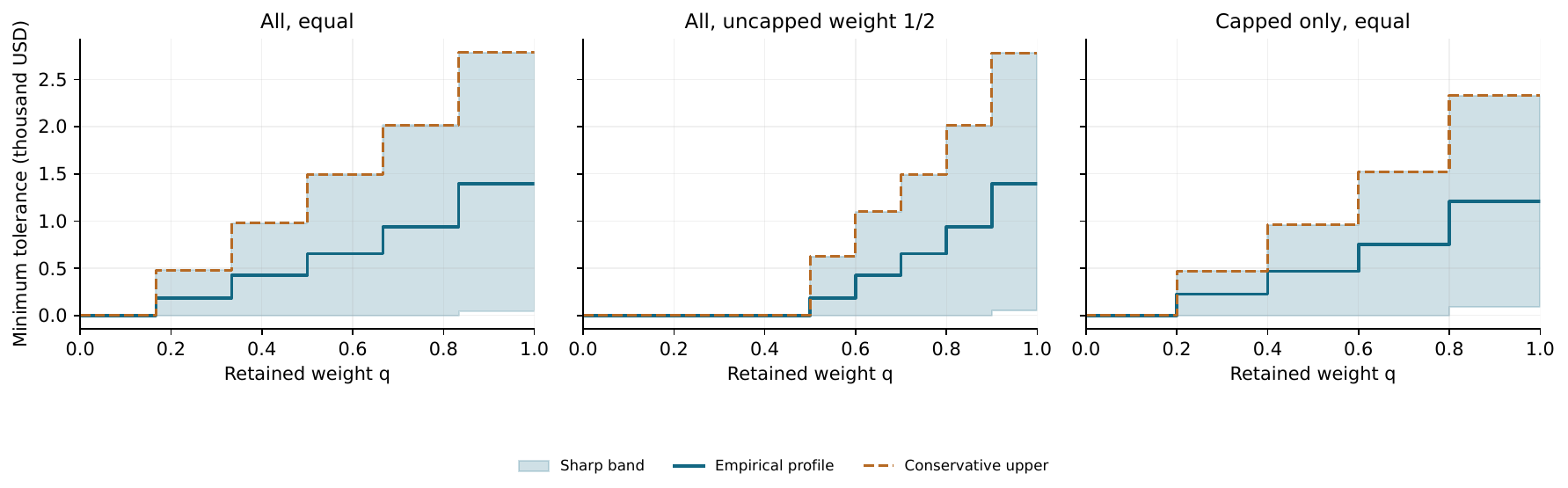}\caption{Reconciliation profiles for three analytical universes: all six definitions with equal weights, all six with the uncapped definition receiving half of total weight, and five capped outcomes with equal weights. Each panel has its own nominal 95\% plug-in maxT calibration. Sharp and conservative upper bounds overlap in these data.}\label{supp:fig:universes}\end{figure}
\begin{figure}[!htbp]\centering
\begin{subfigure}[t]{.49\linewidth}\includegraphics[width=\linewidth]{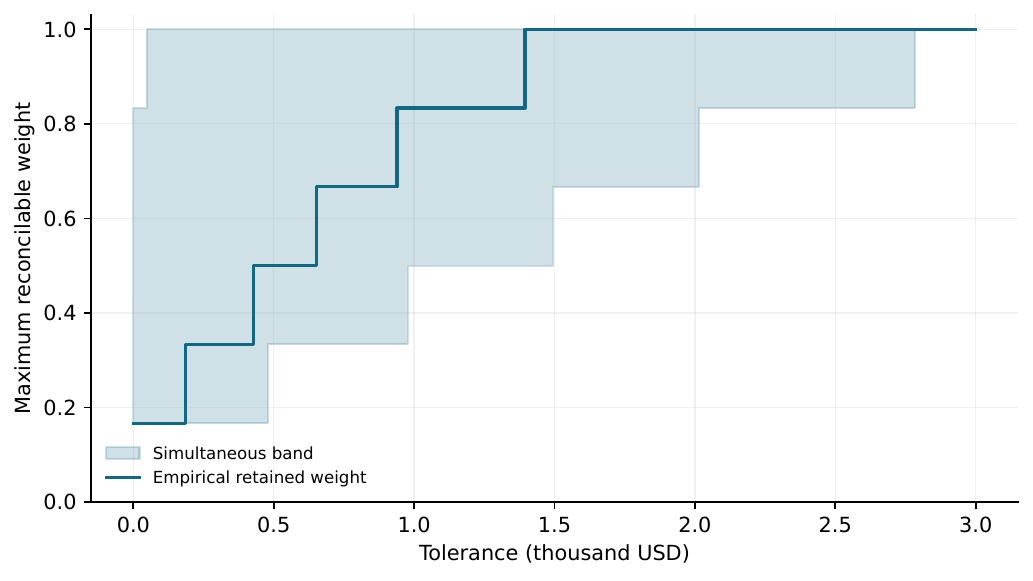}\caption{Inverse profile in the equal-weight full universe.}\label{supp:fig:inv}\end{subfigure}\hfill
\begin{subfigure}[t]{.49\linewidth}\includegraphics[width=\linewidth]{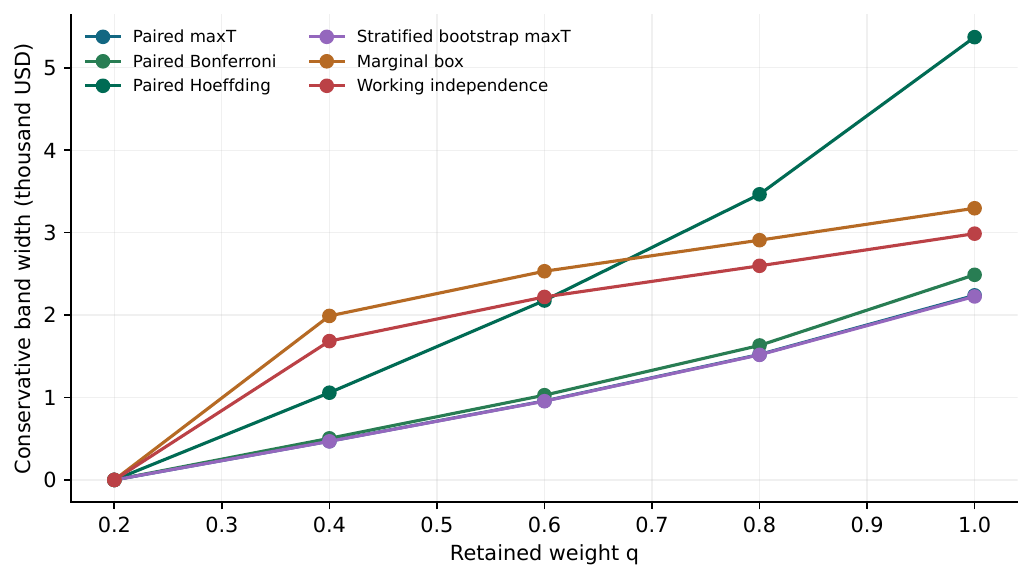}\caption{Widths by calibration in the capped-only universe.}\label{supp:fig:cal}\end{subfigure}
\caption{Tolerance-based interpretation and calibration sensitivity.}\label{supp:fig:realsens}\end{figure}
Figure~\ref{supp:fig:universes} compares different scientific weighting policies, not repeated estimates of a single profile. The uncapped-half scheme necessarily assigns a zero reconciliation width up to half the total weight; this follows from the target definition. Figure~\ref{supp:fig:realsens} shows that at a tolerance of one thousand dollars the primary inverse band is $[0.5,1]$, with estimate $5/6$. The upper endpoint one means the procedure cannot rule out universal agreement at that tolerance; it does not establish universal agreement. The bounded-score construction is much wider because it uses known ranges rather than the observed covariance. The primary Gaussian critical value is 2.55046; the stratified bootstrap value is 2.60228, while the normal Bonferroni value is 2.93520. For the capped-only universe, these values are 2.45678, 2.44281, and 2.80703, respectively. The calibration draws are Monte Carlo approximations, and the modest positive full-range lower endpoint is not invariant to the calibration choice.

\begin{table}[!htbp]\centering\scriptsize\caption{All six definitions, equal weights. Intervals are lower and conservative upper endpoints, in thousands of dollars. M: plug-in paired maxT; B: paired Bonferroni; S: stratified bootstrap maxT; X: marginal box; I: working independence. For M the sharp upper endpoint equals the conservative one to recorded precision.}\label{supp:tab:all_equal}
\resizebox{\linewidth}{!}{\begin{tabular}{lrrrrrr}\toprule $q$&Estimate&M&B&S&X&I\\\midrule$1/6$ & 0.000 & [0.000, 0.000] & [0.000, 0.000] & [0.000, 0.000] & [0.000, 0.000] & [0.000, 0.000] \\
$2/6$ & 0.185 & [0.000, 0.479] & [0.000, 0.517] & [0.000, 0.484] & [0.000, 2.027] & [0.000, 1.740] \\
$3/6$ & 0.427 & [0.000, 0.977] & [0.000, 1.054] & [0.000, 0.987] & [0.000, 2.575] & [0.000, 2.287] \\
$4/6$ & 0.653 & [0.000, 1.493] & [0.000, 1.620] & [0.000, 1.510] & [0.000, 2.956] & [0.000, 2.672] \\
$5/6$ & 0.939 & [0.000, 2.015] & [0.000, 2.177] & [0.000, 2.037] & [0.000, 3.346] & [0.000, 3.069] \\
$6/6$ & 1.395 & [0.047, 2.784] & [0.000, 2.994] & [0.024, 2.813] & [0.000, 3.729] & [0.000, 3.462] \\
\bottomrule\end{tabular}}\end{table}
\begin{table}[!htbp]\centering\scriptsize\caption{All six definitions with half of the total weight assigned to the uncapped outcome. Abbreviations and units as in Table~\ref{supp:tab:all_equal}.}\label{supp:tab:all_weighted}
\resizebox{\linewidth}{!}{\begin{tabular}{lrrrrrr}\toprule $q$&Estimate&M&B&S&X&I\\\midrule$1/10$ & 0.000 & [0.000, 0.000] & [0.000, 0.000] & [0.000, 0.000] & [0.000, 0.000] & [0.000, 0.000] \\
$2/10$ & 0.000 & [0.000, 0.000] & [0.000, 0.000] & [0.000, 0.000] & [0.000, 0.000] & [0.000, 0.000] \\
$3/10$ & 0.000 & [0.000, 0.000] & [0.000, 0.000] & [0.000, 0.000] & [0.000, 0.000] & [0.000, 0.000] \\
$4/10$ & 0.000 & [0.000, 0.000] & [0.000, 0.000] & [0.000, 0.000] & [0.000, 0.000] & [0.000, 0.000] \\
$5/10$ & 0.000 & [0.000, 0.000] & [0.000, 0.000] & [0.000, 0.000] & [0.000, 0.000] & [0.000, 0.000] \\
$6/10$ & 0.185 & [0.000, 0.625] & [0.000, 0.692] & [0.000, 0.635] & [0.000, 3.529] & [0.000, 2.820] \\
$7/10$ & 0.427 & [0.000, 1.099] & [0.000, 1.202] & [0.000, 1.114] & [0.000, 3.622] & [0.000, 2.955] \\
$8/10$ & 0.653 & [0.000, 1.491] & [0.000, 1.620] & [0.000, 1.510] & [0.000, 3.693] & [0.000, 3.077] \\
$9/10$ & 0.939 & [0.000, 2.013] & [0.000, 2.177] & [0.000, 2.037] & [0.000, 3.717] & [0.000, 3.205] \\
$10/10$ & 1.395 & [0.050, 2.781] & [0.000, 2.994] & [0.024, 2.813] & [0.000, 3.729] & [0.000, 3.462] \\
\bottomrule\end{tabular}}\end{table}
\begin{table}[!htbp]\centering\scriptsize\caption{Five capped definitions, equal weights. H denotes the stratified paired Hoeffding construction; other abbreviations as in Table~\ref{supp:tab:all_equal}. The exact upper endpoints for M and H coincide with their conservative counterparts.}\label{supp:tab:capped}
\resizebox{\linewidth}{!}{\begin{tabular}{lrrrrrrr}\toprule $q$&Estimate&M&B&S&X&I&H\\\midrule$1/5$ & 0.000 & [0.000, 0.000] & [0.000, 0.000] & [0.000, 0.000] & [0.000, 0.000] & [0.000, 0.000] & [0.000, 0.000] \\
$2/5$ & 0.226 & [0.000, 0.470] & [0.000, 0.505] & [0.000, 0.468] & [0.000, 1.990] & [0.000, 1.684] & [0.000, 1.058] \\
$3/5$ & 0.468 & [0.000, 0.958] & [0.000, 1.028] & [0.000, 0.955] & [0.000, 2.532] & [0.000, 2.220] & [0.000, 2.176] \\
$4/5$ & 0.754 & [0.000, 1.522] & [0.000, 1.631] & [0.000, 1.517] & [0.000, 2.908] & [0.000, 2.597] & [0.000, 3.465] \\
$5/5$ & 1.210 & [0.090, 2.329] & [0.000, 2.489] & [0.097, 2.323] & [0.000, 3.296] & [0.000, 2.988] & [0.000, 5.372] \\
\bottomrule\end{tabular}}\end{table}
Tables~\ref{supp:tab:all_equal}--\ref{supp:tab:capped} give numerical endpoints for each attainable weight and calibration. These are not simultaneous guarantees over all separately calibrated sensitivity analyses. Working independence remains a diagnostic. The public repository contains the implementation and the notebook under the MIT license: \url{https://github.com/debashisdotchatterjee/Reconciliation_Profile_Stat2026}.

\begin{figure}[!htbp]
\centering\includegraphics[width=.92\linewidth]{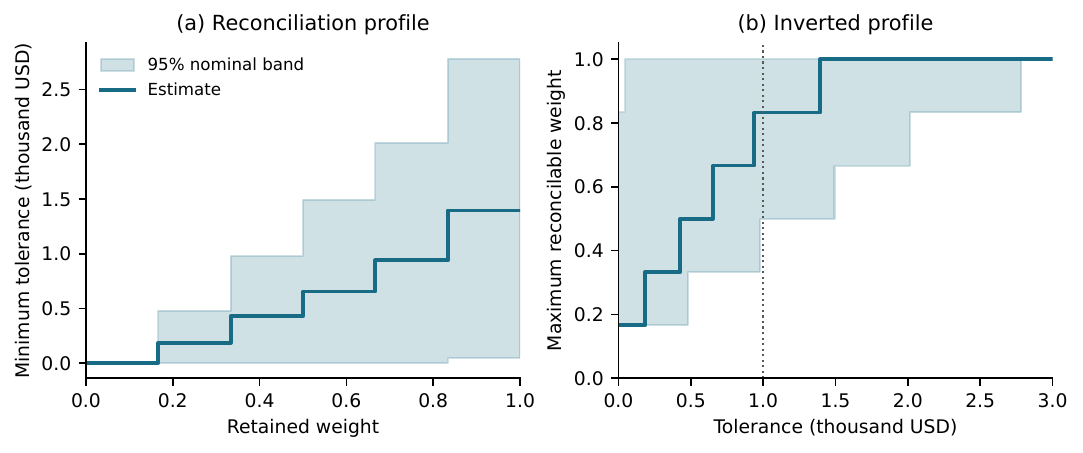}
\caption{Primary equal-weight earnings universe: empirical reconciliation profile with sharp nominal 95\% simultaneous band (a), and inverse maximum reconcilable weight (b). The dotted vertical line indicates a tolerance of one thousand historical US dollars. Both panels represent step functions.}\label{supp:fig:primarycompact}
\end{figure}
Figure~\ref{supp:fig:primarycompact} summarizes the profile and its inversion for the primary universe, complementing the sensitivity plots above.

\section{Reproducibility and run audit}\label{supp:sec:repro}
The research run used seed 20261129, 1,000 replications per scenario, 120,000 Gaussian calibration draws, 10,000 bootstrap draws per earnings universe, and 150 exact-band replications per scenario. The run-status file marks simulation, numerical audit, exact-band simulation, and real-data analysis complete. The replication-level summaries contain 144,000 distinct scenario--method--replication rows (24,000 data-generating replications evaluated by six methods), all with 1,000 replicates per scenario--method pair. Reaggregation reproduces the submitted coverage summaries exactly and mean widths within $5.6\times10^{-17}$. The 3,600 exact-band rows represent 900 replications evaluated at four weights. The accompanying \texttt{audit/} directory retains summary CSV files, calibration records, configuration, and provenance; the larger replication-level archive is retained separately by the author. The numerical results and original plots are generated by the repository code, with the two compact main-text figures regenerated directly from the exported summaries.

\clearpage
\bibliographystyle{biometrika}
\bibliography{reference}
\end{document}